\documentclass[12pt,a4paper]{article}
\usepackage{epsfig}
\def\eq#1{{Eq.~(\ref{#1})}}

\newcommand{\beq}{\begin{equation}}
\newcommand{\eeq}{\end{equation}}
\newcommand{\beqar}[1]{\begin{eqnarray}\label{#1}}
\newcommand{\eeqar}{\end{eqnarray}}
\def\arnps#1#2#3{  {\it Ann. Rev. Nucl. Part. Sci. }{\bf #1}:#2 (#3)}
\def\npb#1#2#3{    {\it Nucl. Phys. }{\bf B#1}:#2 (#3)}
\def\plb#1#2#3{    {\it Phys. Lett. }{\bf B#1}:#2 (#3)}
\def\prd#1#2#3{    {\it Phys. Rev. }{\bf D#1}:#2 (#3)}
\def\prep#1#2#3{   {\it Phys. Rep. }{\bf #1}:#2 (#3)}

\def\zpc#1#2#3{    {\it Z. Phys. }{\bf C#1}:#2 (#3)}

\def\sjnp#1#2#3{   {\it Sov. J. Nucl. Phys. }{\bf #1}:#2 (#3)}

\def\jetpl#1#2#3{  {\it JETP Lett. }{\bf #1}:#2 (#3)}
\def\epjc#1#2#3{   {\it Eur. Phys. J.}{\bf C#1}:#2 (#3)}

\begin{document}
\title {{\bf Higher twists and maxima for DIS }\\
{\bf   on nuclei in high density QCD region}}
\author{
{\bf
E.~Gotsman\thanks{e-mail: gotsman@post.tau.ac.il}~$\, ^a$,
\quad  E.~Levin\thanks{e-mail: leving@post.tau.ac.il,
 elevin@quark.phy.bnl.gov}~$\, ^{a,b}$,
\quad U.~Maor\thanks{e-mail: maor@post.tau.ac.il}~$\, ^a$,
} \\
{\bf
L.~McLerran\thanks{e-mail: mclerran@bnl.gov}~$\, ^b$,
 \quad and \quad
K.~Tuchin\thanks{e-mail: tuchin@post.tau.ac.il}~$\, ^a$,
} \\[10mm]
{\it\normalsize $^a$HEP Department}\\
 {\it\normalsize School of Physics and Astronomy,}\\
 {\it\normalsize Raymond and Beverly Sackler Faculty of Exact Science,}\\
 {\it\normalsize Tel-Aviv University, Ramat Aviv, 69978, Israel}\\[0.5cm]
{\it\normalsize $^b$ Physics Department,}\\
{\it\normalsize Brookhaven National Laboratory,}\\
 {\it\normalsize Upton, NY 11973-5000, USA}
}

\date{July 2000}
\maketitle
\thispagestyle{empty}

\begin{abstract}
   We show that the ratio of different structure functions have a
maximum which depends on $x_B$ and $A$. We argue that these maxima are
proportional to the saturation scale. The analysis of leading and higher
twist contributions for different observables is given
 with the aim of determining the
kinematic region where high parton density effects could be seen
experimentally. 

 \end{abstract}
\thispagestyle{empty}
\begin{flushright}
\vspace{-17cm}
TAUP-2638 - 2000\\
BNL-NT-00/19\\
\today
\end{flushright}

\newpage
\setcounter{page}{1}



\section{Introduction}
It was noted many years ago that at fixed $Q^2$ but low $x_B$
(high
energies)  we are dealing with a parton system in DIS in  a  kinematical
region  where the coupling constant is small while the interaction between
constituents is large\cite{GLR}. In such a parton system we have 
competition between emission of partons (gluons) and their annihilation,
which leads to an equilibrium state with a definite value of the gluon
density\cite{GLR}\cite{MUQI}\cite{MCVE}. At any  value of small $x_B$
there
is a value of $Q^2= Q^2_s(x_B)$ at which the parton density 
reaches a sufficiently high 
value that the interactions (annihilations) between partons become
so strong that they 
  diminish the number of partons. The new scale in DIS at which
this phenomenon occurs we shall call the saturation scale. This scale
depends
on the energy of the colliding system as well as on the atomic numbers
of colliding nuclei, and can be evaluated from the
condition
that the packing factor ($f_{pack}$) of partons in  a parton cascade is
equal to
unity\cite{GLR,MUQI,MCVE} :
\begin{equation} \label{EQ1}
f_{pack} \, =\,\,\,\frac{3\,\pi^2 \alpha_S}{2
Q^2_s(x_B;A)}\,\times\,
\frac{x_BG_A(x_B,Q^2_s(x_B;A))}{\pi\, R^2_A}\,\,=\,\,1\quad ;
\end{equation}
and $r^2_{saturation} = 1/Q^2_s(x_B)$.

The high parton density effects appear as  nonlinear
corrections to the DGLAP evolution equation.
A few theoretical approaches have been  developed which take
into account non-linear corrections to the DGLAP evolution. One of them
is based on the pQCD approach which has been used at the limit of the
region where it is valid
\cite{GLR,MUQI,HDPQCD}. Finally, Yu.~Kovchegov \cite{KOV}
and Ia.~Balitsky \cite{BALITSKY} have derived an evolution equation for
the dipole wave function
with GLR-type corrections\cite{GLR} which describes the parton system in
the full kinematic region. The initial condition for this
equation is a Glauber-like distribution which must be fixed at not too
small values of $x_B$.
 The saturation scale which appears in this equation has been
studied in Refs.\cite{LT}. The same scale appears in a different approach
based on the effective Lagrangian for semiclassical gluonic fields,
emission
and the interactions which are responsible for high gluon density system
\cite{MCVE, HDSCGF}. Despite the  quite different techniques that have
been
incorporated in these approaches, they describe the same physics just
approaching it from a different value of the parton density: pQCD from
the low
density side while the effective Lagrangian approach is from the high
density side. 

Approaching
the saturation scale  the total cross section saturates at a geometrical
size of the
target. Using the language of the operator product expansion  we can say
that the reason for this breakdown is
that it is no longer a valid approximation to neglect higher twist
operators in comparison with the twist-2  operator. Thus,  to
estimate
the  kinematical region for which  shadowing corrections , i.e.  
non-linear effects at large energies  become
important, one has to solve the non-linear evolution equation.

 This problem has not yet been solved, to do so we suggest the
following approach. Namely,
we propose estimating the size of the different twist contributions to the
observed quantities and find the kinematical region where they  become
equally important. This will aid in determining the region where the
linear evolution breaks down. Such an approach has been developed in Ref.
\cite{BARTELS} for the DIS on a nucleon in the framework of Golec-Biernat
and Wusthoff model \cite{WUST}.

It should be stressed that in spite of the fact that at present
 there does not appear to be any direct evidence of saturation
 in the experimental data at HERA, there are some hints
 of high parton density effects in the same data \cite{HERADD}. One of the
goals of this paper is  to provide experimentalists with new observables
or/and with new ways of  analyzing the data  which will enable one to
explore 
high density QCD effects in  data from nuclear targets.

The paper is  organized as follows:
We  start by arguing that the eikonal approach is a reasonable model,
since it is a first
iteration of the exact evolution equation. We then derive, making an
approximation, the analytical expression for the contribution of any twist
to the longitudinal and transverse structure functions, as well as
to the longitudinal and transverse diffractive structure
functions. We show that   to estimate the saturation scale
it is helpful to consider
the ratios of the various structure functions. 
(Sec.~2). In Sec.~3 we discuss numerical estimations with formulae of
Sec.~2.
 Finally, in Sec.~4 we conclude by proposing how one can measure different
twists at eRHIC.

\section{The Model}

\subsection{Eikonal Approach}
Complete description of the eikonal (or Mueller-Glauber) approach to
shadowing
corrections  was given in Refs.~\cite{DIPOLE,MU94,NN,DIX,Nucleon}.
We recall the
main
points.
\begin{enumerate}
\item\quad We work in the target rest frame where the whole  evolution
occurs in the virtual photon wave function. This wave function consists of
colour dipoles which
emerged as a result of numerous splittings of gluons (i.e.\ quarks and
anti-quarks in $N_c\gg 1$). In the leading $\log(1/x_B)$ approximation a
long
time elapses after the photon wave
function is formed, until every dipole scatters off the
nucleus. The factorization of the two processes --- radiation of the
dipoles
and
their interaction with the target is the basis of the 
dipole picture of DIS\cite{DIPOLE,MU94}. In this picture the
scattering amplitude is
diagonal with respect to the size of the dipole $r_t$ and the fraction of
initial energy $z$ carried by quark \cite{MU94} .
Let $P^{\gamma^*}(z,r_t;Q^2)$ be the probability to find a quark -antiquark
pair  inside a virtual photon of virtuality $Q^2$ and
$\hat\sigma(x_B,r_t)$ be
the total cross
section for the dipole-nucleon interaction.  The total cross section
for the interaction of the virtual photon with the nucleus is given by
\cite{DIPOLE,MU94,NN,DIX}
\beq\label{DIP1}
\sigma_{tot}(\gamma^*p)(x_B,Q^2)=
\int\, dz\int\, d^2r_t\, P^{\gamma^*}(z,r_t;Q^2)\hat\sigma(x_B,r_t)
\eeq
\begin{eqnarray}
P^{\gamma^*}(z,r_t;Q^2)&=&\frac{\alpha_{em}N_c}{2\pi^2}\sum_f Z_f^2
\{ |\Psi_T|^2+|\Psi_L|^2\}\\
|\Psi_T(z,r_t;Q^2)|^2&=&(z^2+(1-z)^2)a^2K_1^2(ar_t)+m_f^2\,
K_0^2(ar_t)\\
|\Psi_L(z,r_t;Q^2)|^2&=&4Q^2z^2(1-z)^2K_0^2(ar_t) 
\label{DIP2}
\end{eqnarray}
where $a^2=z(1-z)Q^2 +m_f^2$.

\item\quad The optical theorem in the impact parameter $b_t$
representation reads 
\beq\label{IMP}
\hat\sigma(x_B,r_t)=2\int\, d^2 b_t \,\mathrm{Im}\,
a^{el}_{dipole}(x_B,r_t,b_t)
\eeq
where $a^{el}(x_B,b_t)$ is a partial elastic
amplitude at fixed $b_t$ with a given Bjorken scaling variable  $x_B$.

\item\quad In all approaches to high energy scattering processes, it is
assumed that the 
dipole -- nucleus amplitude is purely imaginary at high energy.
$s$-channel unitarity then leads to the following equation:
\beq\label{CONSTR}
a^{el}_{dipole}(x_B,t_t;b_t) =i\left(
1-e^{-\frac{\Omega(x_B,r_t;b_t)}{2}} \right) \eeq
In the Glauber-Mueller  model the opacity
$\Omega(x_B,r_t;b_t)$ represents the exchange of one hard Pomeron. It
is defined  as
\beq\label{OMEGA}
\Omega(x_B,r_t;b_t)=\kappa(x_B,r_t)\: S(b_t)\:\pi R_A^2
=\frac{\pi^2\alpha_S}{3}\,
r_t^2\, x_BG(x_B,\frac{4}{r_t^2})\, S(b_t)\quad ,
\eeq
so that for small  $\Omega(x_B,r_t;b_t)$ \eq{IMP}  yields the DGLAP
evolution equation
(in the double $\ln(1/x_B)\ln(Q^2)$) approximation. We assume that this
relation is valid in the whole kinematical region.

\item\quad We assume, for simplicity, a Gaussian form for the profile
function
\beq\label{PROFILE}
S(b_t)=\frac{1}{\pi R_A^2}e^{-\frac{b_t^2}{R_A^2}}
\eeq
where $R_A$ is a target radius.

\item Diffraction of the photon on the target can be thought of as an
elastic scattering of each dipole off the nucleus \cite{MLK}. So instead
of
$\mathrm{Im}\, a^{el}_{dipole}(x_B,r_t,b_t)$ we substitute
\beq\label{aDIF}
|a^{el}_{dipole}(x_B,r_t,b_t)|^2=
\left(
1-e^{-\frac{\Omega(x_B,r_t;b_t)}{2}} \right)^2
\eeq
to get $\sigma^D(\gamma^*A)$.
\end{enumerate}

It was shown in Ref.~\cite{Nucleon,NUCLEUS, BRAZ} that the
eikonal
approximation leads to a  reasonable
theoretical as well as experimental approximation in the region
of not too small $x_B$. We will use this model to
calculate contributions of different twists to the total and
diffractive cross  sections. This needs additional justification
which are based on the following arguments.
\begin{enumerate}
\item\quad When the shadowing corrections  are small ( at large values of
$Q^2$ ) the eikonal
approximation gives
the correct matching to the  DGLAP evolution;

\item\quad The eikonal approximation gives a natural explanation and an
estimate of the saturation scale $Q_s(x_B,A)$ \cite{GLR,MUQI,MCVE};

\item It is a good approximation\cite{BRAZ}   to the non-linear evolution
equation for
the HERA kinematical region at $Q^2 \geq 5\,\,GeV^2 $  where it describes
available experimental data  well.

\item\quad It provides the  correct anomalous dimensions for all higher
twist contributions in the limit $N_c\gg 1$ and 
$\alpha_S(Q^2)\ln(1/x_B)\gg 1$\cite{ANOMAL};

\item\quad Our approach includes the impact parameter behaviour of the
dipole-target scattering amplitude and, therefore, it can be used for
different target , including nuclei;

\item\quad The eikonal approach is very similar to the saturation
model of Ref.~\cite{WUST} which describes the  experimental data at
HERA very well.
The main difference between these models is that eikonal satisfies
the DGLAP evolution equation in the limit of large $Q^2$, while the
Golec-Biernat and Wusthoff model does not do so (for the complete
discussion of this issue see Ref.~\cite{Nucleon} ).
\end{enumerate}

This simple  eikonal model is a reasonable starting point for
understanding how the shadowing corrections are turned on. Moreover, as
was shown in Ref.~\cite{KOV} using the Glauber approach for a nuclear
target
can be justified, there is a possibility that it will also  be successful 
for the case
of a nucleon  (see Ref. \cite{BRAZ1}).

\subsection{Analytical expansion for $m_f^2=0$.}
Collecting formulae of the previous subsection we arrive at
the following expressions (for three flavours):
\begin{eqnarray}
\sigma_{tot}(\gamma^*A)(x_B,Q^2)&=&2\,\frac{\alpha_{em}}{\pi^2} \int\,
dz\int\, d^2r_t\int\, d^2b_t\; 
\{|\Psi_T|^2+|\Psi_L|^2\}\;
\left(1-e^{-\frac{\Omega(x_B,r_t;b_t)}{2}} \right)\label{TOT}\\
\sigma_{tot}^D(\gamma^*A)(x_B,Q^2)&=&\frac{\alpha_{em}}{\pi^2} \int\,
dz\int\, d^2r_t\int\, d^2b_t\; 
\{|\Psi_T|^2+|\Psi_L|^2\}\;
\left(1-e^{-\frac{\Omega(x_B,r_t;b_t)}{2}} \right)^2\label{DIF}
\end{eqnarray}

We introduce new  dimensionless integration variables which simplify the
calculation
according to
\beq t^2= r_t^2\, Q^2;\quad
\xi =e^{-\frac{b_t^2}{R_A^2}};\quad
 y^2=z(1-z)\quad .
\eeq

In the limit $m_f^2=0$ \eq{TOT} reads:
\begin{eqnarray}
\sigma_L(\gamma^*A)&=&32R_A^2\alpha_{em}\int_0^{1/2}\,dy\int_0^\infty\,
dt^2\int_0^1\, \frac{d\xi}{\xi}\, \frac{y^5}{\sqrt{1-4y^2}}\,
K_0^2(yt)\left(1-e^{-\frac{\Omega}{2}} \right) \\
\sigma_T(\gamma^*A)&=&8R_A^2\alpha_{em}\int_0^{1/2}\,dy\int_0^\infty\,
dt^2\int_0^1\, \frac{d\xi}{\xi}\, \frac{y^3(1-y^2)}{\sqrt{1-4y^2}}\,
K_1^2(yt)\left(1-e^{-\frac{\Omega}{2}} \right)\\
\sigma_{tot}(\gamma^*A)&=& \sigma_L(\gamma^*A)+ \sigma_T(\gamma^*A)
\end{eqnarray}
and a similar expression for \eq{DIF}.

Most of the dipoles in the virtual photon wave function are of size
$r_t^2=4/Q^2$. We assume that \emph{all} dipoles are of  this size.
This assumption will not affect our calculations of the twist
contributions
to the cross sections, as it only gives  logarithmic  $Q^2$ corrections
at small $Q^2$ and is valid in  the leading $\ln Q^2$
approximation. 
Thus, \eq{OMEGA} reads
\beq\label{APPROX}
\frac{\Omega(x_B,t;\xi;Q^2)}{2}\approx \frac{t^2\xi}{Q^2}\, C(x_B,Q^2)
\eeq 
where we have defined\footnote{In our model $C(x_B,Q^2)$ is a saturation
scale. In general, the saturation scale has to follow from the solution of
the complete non-linear evolution equation as we have already mentioned.}
\beq\label{C}
C(x_B,Q^2)\equiv\frac{A\pi\alpha_S\, x_BG(x_B,Q^2)}{6R_A^2}
\eeq
In this massless approximation integrals of \eq{TOT} and \eq{DIF} can
be expressed in terms of special functions.

\begin{figure}\begin{center}
\epsfig{file=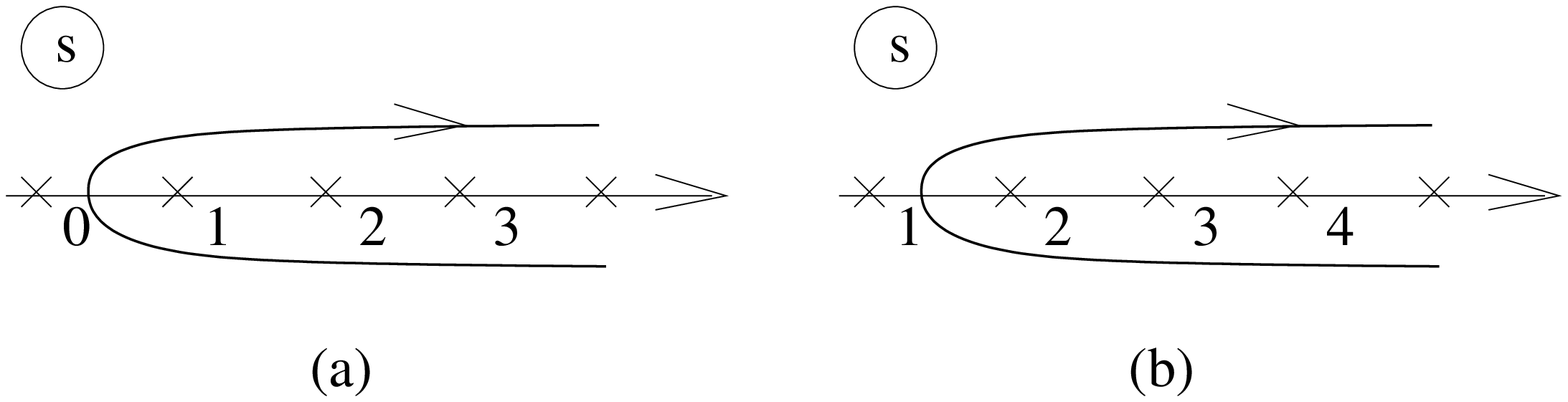, width=14cm,height=3.7cm}
\caption{Integration contours for the Meijer function in \eq{LONG} and 
\eq{TRANS}.}
\label{FIG:CONTOURS}
\end{center}\end{figure}

First, integration over $\xi$ gives:
\beq
\int_0^1\,\frac{d\xi}{\xi}
\left(1-e^{-\frac{t^2C}{Q^2}\xi} \right)
=\ln\left(\frac{t^2C}{Q^2}
\right)+\gamma+E_1 \left(\frac{t^2C}{Q^2}\right)
=\sum_{k=1}^\infty\frac{(-1)^{k+1}}{k\: k!}\left(\frac{t^2C}{Q^2}\right)^k
\eeq
where $E_1$ is an exponential integral and $\gamma$ is the Euler constant.
We now consider  the longitudinal part of the cross section.
Integrating over $t$ and summing over $k$ one obtains
\begin{eqnarray}
\sigma_L(\gamma^*A)&=&\sqrt{\pi}\alpha_{em}R_A^2\int_0^1\,
\frac{dy\, y^5}{\sqrt{1-y^2}}\sum_{k=1}^\infty\,
\frac{(-1)^{k+1}}{k\: k!}\left(\frac{C}{Q^2}\right)^k
\frac{4^k\Gamma(1+k)^3(y^2)^{-1-k}}{\Gamma(\frac{3}{2}+k)}\nonumber\\
&=& \frac{16C}{3Q^2}\alpha_{em}R_A^2\int_0^1\,
\frac{dy\, y}{\sqrt{1-y^2}}\;
{_3}F_1(1,1,2;\frac{5}{2};-\frac{4C}{Q^2y^2})\nonumber\\
&=& 2\alpha_{em}\pi R_A^2\frac{C}{Q^2}
G^{3\, 2}_{3\, 4}
\left(\frac{Q^2}{4C}\Bigg|{0,1;\frac{5}{2}\atop 1,1,2;-\frac{1}{2}}\right)
\label{LONG}
\end{eqnarray}
where we have used the definition of the generalized hypergeometric
function
${_3}F_1$ and expressed the integral over $y$ as a Meijer function
$G^{3\, 2}_{3\, 4}$ (see Ref.\cite{BATEMAN}).
This Meijer function is defined by the following
integration
\beq
G^{3\, 2}_{3\, 4}
\left(\frac{Q^2}{4C}\Bigg|{a_1,a_2;a_3\atop
                                      b_1,b_2,b_3;b_4}\right)
=\frac{1}{2\pi i}\int_L ds \left(\frac{Q^2}{4C}\right)^s
\frac{\Gamma(1-a_1+s)\Gamma(1-a_2+s)\Gamma(b_1-s)\Gamma(b_2-s)\Gamma(b_3-s)}
{\Gamma(a_3-s)\Gamma(1-b_4+s)}
\eeq
where the integration contour $L$ runs from $+\infty$ to $+\infty$
and encloses all poles of functions $\Gamma(b_i-s)$, $i=1,2,3$ in the
negative direction, but does not encloses poles of functions
$\Gamma(1-a_j+s)$, $j=1,2$.
The contour for the Meijer function in \eq{LONG} is shown in
Fig.\ref{FIG:CONTOURS}(a).

In the case of the transverse cross section we repeat the by now familiar
procedure to obtain the following result:
\begin{eqnarray}
\sigma_T(\gamma^*A)&=&4\alpha_{em}\pi R_A^2\frac{C}{Q^2}
\left[ G^{3\, 2}_{3\, 4}
\left(\frac{Q^2}{4C}\Bigg|{1,1;\frac{5}{2}\atop 1,1,1;\frac{1}{2}}\right)
-\frac{1}{2}
G^{3\, 2}_{3\, 4}
\left(\frac{Q^2}{4C}\Bigg|{0,1;\frac{5}{2}\atop 1,1,1;-\frac{1}{2}}\right)
\right.
\nonumber\\[3mm]
&&\left. +\frac{4C}{Q^2}\left\{\frac{1}{4}
G^{3\, 2}_{3\, 4}
\left(\frac{Q^2}{4C}\Bigg|{1,1;\frac{7}{2}\atop 2,2,2;\frac{1}{2}}\right)
-\frac{1}{2}
G^{3\, 2}_{3\, 4}
\left(\frac{Q^2}{4C}\Bigg|{1,2;\frac{7}{2}\atop 2,2,2;\frac{3}{2}}\right)
\right\}\right]\label{TRANS}
\end{eqnarray}
The contour for the first two functions is shown in
Fig.\ref{FIG:CONTOURS}(a) and for the last two in
Fig.\ref{FIG:CONTOURS}(b).

The simplest way to calculate $\sigma(\gamma^*A)^D_L$ and
$\sigma(\gamma^*A)^D_T$ is to note that using
 \eq{CONSTR} and \eq{aDIF}
$$
\sigma_{el}(\kappa)=\int\, d^2b_t \left(1-e^{-\frac{\Omega}{2}}\right)^2=
2\int\, d^2b_t \left(1-e^{-\frac{\Omega}{2}}\right)-
\int\, d^2b_t \left(1-e^{-\Omega}\right) =
\hat\sigma(\kappa)-\frac{1}{2}\hat\sigma(2\kappa)
$$
Using \eq{LONG} and \eq{TRANS} one gets
\beq\label{DIFFRACT}
\sigma(\gamma^*A)^D_{T,L}(C/Q^2)=
\sigma(\gamma^*A)_{T,L}(C/Q^2)
-\frac{1}{2}\sigma(\gamma^*A)_{T,L}(2C/Q^2)
\eeq

Formulae \eq{LONG}, \eq{TRANS} and \eq{DIFFRACT} make it possible
to estimate the contribution of any twist to the particular cross section.
To this  end we should close the contours of Fig.\ref{FIG:CONTOURS} to
the left over a particular number of poles. Each of these poles represents
the contribution of a particular twist. Below we show some results for
twist-2
and twist-4 contributions.\footnote{
Closing the contours to the right
provides a small $Q^2$ series which, however is of little interest, since
we assumed that $m_f^2=0$.}
\begin{eqnarray}
\sigma_L^{\tau=2}&=&2\alpha_{em}\pi R_A^2\frac{C}{Q^2}\frac{8}{3\pi}
\quad ,\\
\sigma_T^{\tau=2}&=&4\alpha_{em}\pi R_A^2\frac{C}{Q^2}
\frac{14+12(\gamma+\ln(Q^2/4C))}{9\pi}\quad ,\\
\sigma_L^{D\,\tau=2}&=&0\quad ,\\
\sigma_T^{D\,\tau=2}&=&4\alpha_{em}\pi R_A^2\frac{C}{Q^2}\ln 2
\quad ,\\[0.5cm]
\sigma_L^{\tau=4}&=&-2\alpha_{em}\pi R_A^2\left(\frac{C}{Q^2}\right)^2
\frac{32(17+30\gamma+30\ln(Q^2/4C))}{225\pi}\quad ,\\
\sigma_T^{\tau=4}&=&4\alpha_{em}\pi R_A^2\left(\frac{C}{Q^2}\right)^2
\frac{24}{15\pi}\quad .
\end{eqnarray}

\subsection{Ratios of cross sections}

 As we now have analytical formulae for different twists we can
estimate in which kinematical region twist-2, say,
equals twist-4, and therefore  find where the twist expansion breaks down.
We will do this in the following section. The estimation we make is
not accurate for two reasons. First, the equality of two first twists is a
sufficient but not  necessary condition  for the breaking down of the
twist
expansion. This argument will be illustrated in the next section.
 Second,  all models describing the shadowing corrections  are written in
the
leading logarithmic 
approximation. In order to estimate the reliability of these 
models one has to perform calculation of the corrections to the  
leading order. This has not been done yet. In
Ref.~\cite{Nucleon,BRAZ} we overcame this problem by noting that in
the eikonal approximation corrections to the
leading log factorize out of the formulae for the cross sections. This
makes it possible to describe the experimental data quite well by
introducing some universal factor. 
However, we would like to avoid introducing factors of unclear
theoretical origin.

One can overcome these difficulties by studying  the
ratios of the cross sections instead of the cross sections themselves.
In this paper we will consider two ratios $\sigma_L/\sigma_T$ and
$\sigma_L^D/\sigma_T^D$. Both of these ratios have a remarkable property
as
a function of photon virtuality $Q^2$. At small $Q^2$ the ratios vanish
since the real photon is transverse. At large $Q^2$ they vanish as well,
as  implied by pQCD . This leads us to suggest that both ratios have
maximum.
We would expect that this maximum occurs at the same value of $Q^2_{max}$
indicating that there exists a universal scale $Q_s^2\propto Q^2_{max}$ 
which we call the "saturation
scale".  If in the kinematical region that we are interested in we get
different
maxima for different ratios, then  we can  conclude that the
Shadowing Corrections are small in this specific kinematical region. The
maximum is  strongly influenced by the  confinement forces at the scale
$m_f^2$.

In practice it is  difficult to separate the contributions of unitarity
and confinement forces. The manifestation of this is that
$\sigma_L$ and $\sigma_L^D$ are logarithmically
divergent in the massless limit as $Q^2\rightarrow 0$.
We, however,  expect that at large enough energies (small $x_B$)
the unitarity forces become important at much larger $Q^2$ then the
confinement ones. This should appear as a maximum at virtualities much
larger
than the quark mass squared, which has been introduced to model
confinement forces.

\section{Numerical estimations}

We now present the results of numerical calculations. We
performed the calculation using GRV'94 parameterization for the
gluon structure function $x_BG(x_B,Q^2)$. The reason for choosing
this particular parameterization is that we hope that non-linear
corrections to the DGLAP evolution equation which  were, perhaps,
seen at HERA are not obscured by it. It is well known that all
existent experimental data can be fitted more or less well with a linear
parameterization. It seems  senseless to look for shadowing
corrections using such parameterization. An additional reason for
using GRV'94 is that it enables us to perform calculations at values of
$Q^2$ as low as\footnote{
See full discussion in Ref.~\cite{Nucleon}.}
$0.4$ GeV$^2$.

However, GRV'94 has a serious drawback. There exists a domain in
$Q^2$ where it gives the anomalous dimension of $x_BG(x_B,Q^2)$
larger than 1. This manifests itself as a spurious minimum of
$1/\kappa(x_B,4/Q^2)$ (see \eq{OMEGA}) as a function of $Q^2$
at some $\tilde Q^2(x_B)$.
This minimum threatens to spoil the calculation of extremum of
ratios discussed in the previous section. To get rid of it we use
the following gluon structure function
\begin{eqnarray}
x_BG(x_B,Q^2)&=&x_BG^{GRV}(x_B,Q^2)\:\theta(Q^2-\tilde Q^2(x_B))
\nonumber\\
&&+
\frac{Q^2}{\tilde Q^2(x_B)}\: x_BG^{GRV}(x_B,\tilde Q^2(x_B))
\:\theta(\tilde Q^2(x_B)-Q^2) \label{trick}
\end{eqnarray}
where $\theta$ is a step function. \eq{trick} is a rough way of taking
into account the  shadowing corrections in the gluon channel which have
not
been calculated systematically  in our approach, namely, \eq{trick} leads
to a DGLAP gluon structure function for $Q^2$ larger than the saturation
scale $ \tilde{Q}^2(x_B)$ while it gives a  saturated  gluon structure
function for $Q^2$ smaller than this scale.

The results are conveniently represented in
terms of various structure functions instead of
corresponding cross sections. They are defined  by the equation
$$
F_2(x_B,Q^2)=\frac{Q^2}{4\alpha_{em}\pi^2}\sigma(\gamma^*A)
$$

Note, that all the equations of our model are written in the leading 
$\ln(1/x_B)$ approximation\footnote{
And leading $\ln Q^2$ as well, see explanations to the \eq{APPROX}.}. 
This means that they are valid in the
kinematical region where the condition $\alpha_S(Q^2)\ln(1/x_B)
\ge 1$ holds. Hence, we are working at the boundary of this kinematical
region.  So
the question concerning  corrections to the eikonal approach arises.
These corrections are large but, as we have discussed, they cancel in the
ratios since they mostly affect the normalization of the calculated
structure functions. We can  take these corrections into account by
modeling
the anomalous dimension as described in Ref.~\cite{BRAZ}. This effectively 
leads to the factor $\approx 0.5$ in front of \eq{LONG} and \eq{TRANS}.
 To compare our calculation of the structure functions 
with experimental data one should divide all the values by factor 2.

\begin{figure}
\begin{flushleft}
\begin{tabular}{ccc}
\multicolumn{3}{c}{\rule[-3mm]{0mm}{4mm} $ F_L(Q^2)$}\\
$x_B=10^{-2}$&$x_B=5\cdot 10^{-3}$& $x_B=10^{-3}$\\[-10mm]
\epsfig{file=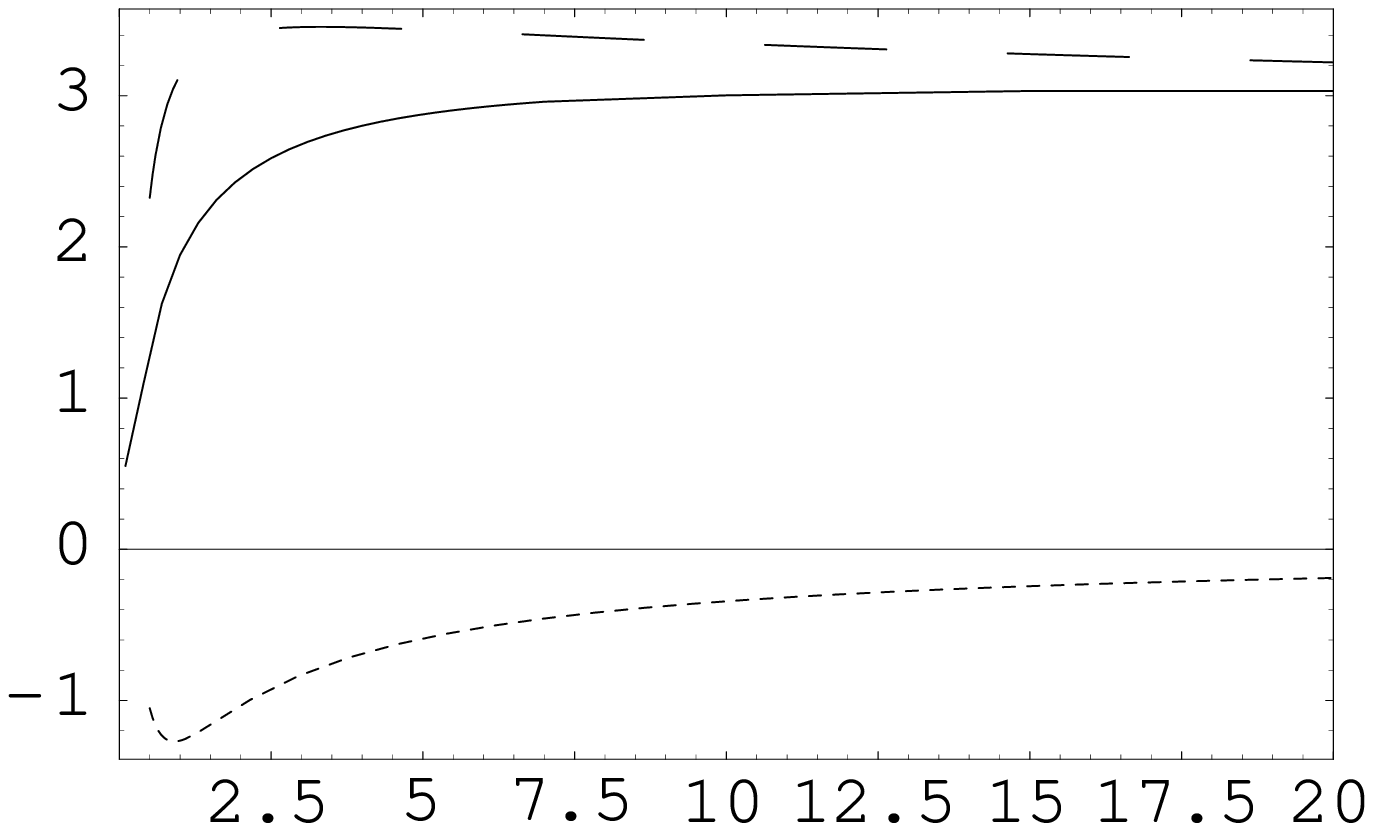,width=50mm,height=50mm}
& \epsfig{file=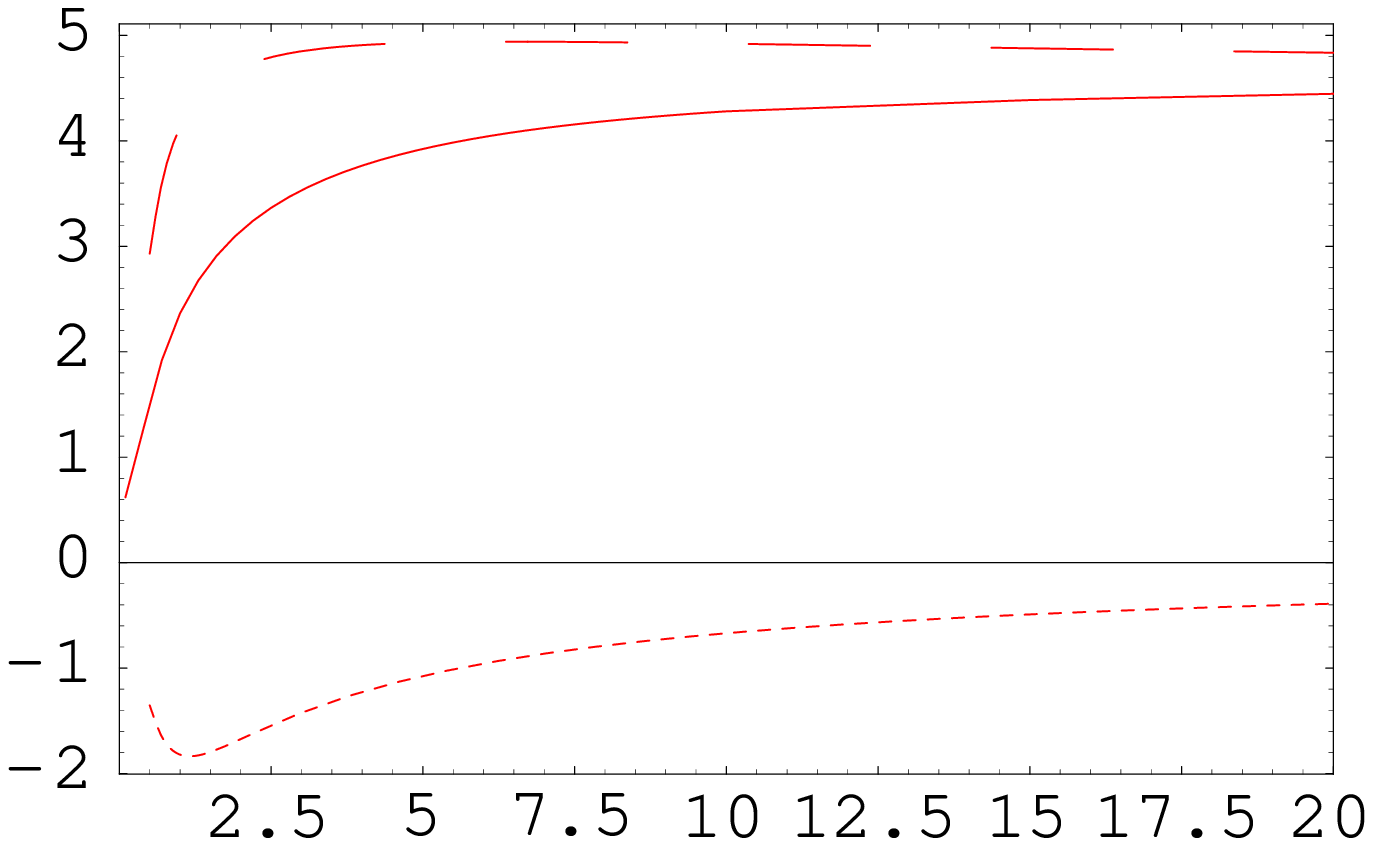,width=50mm,height=50mm}&
\epsfig{file=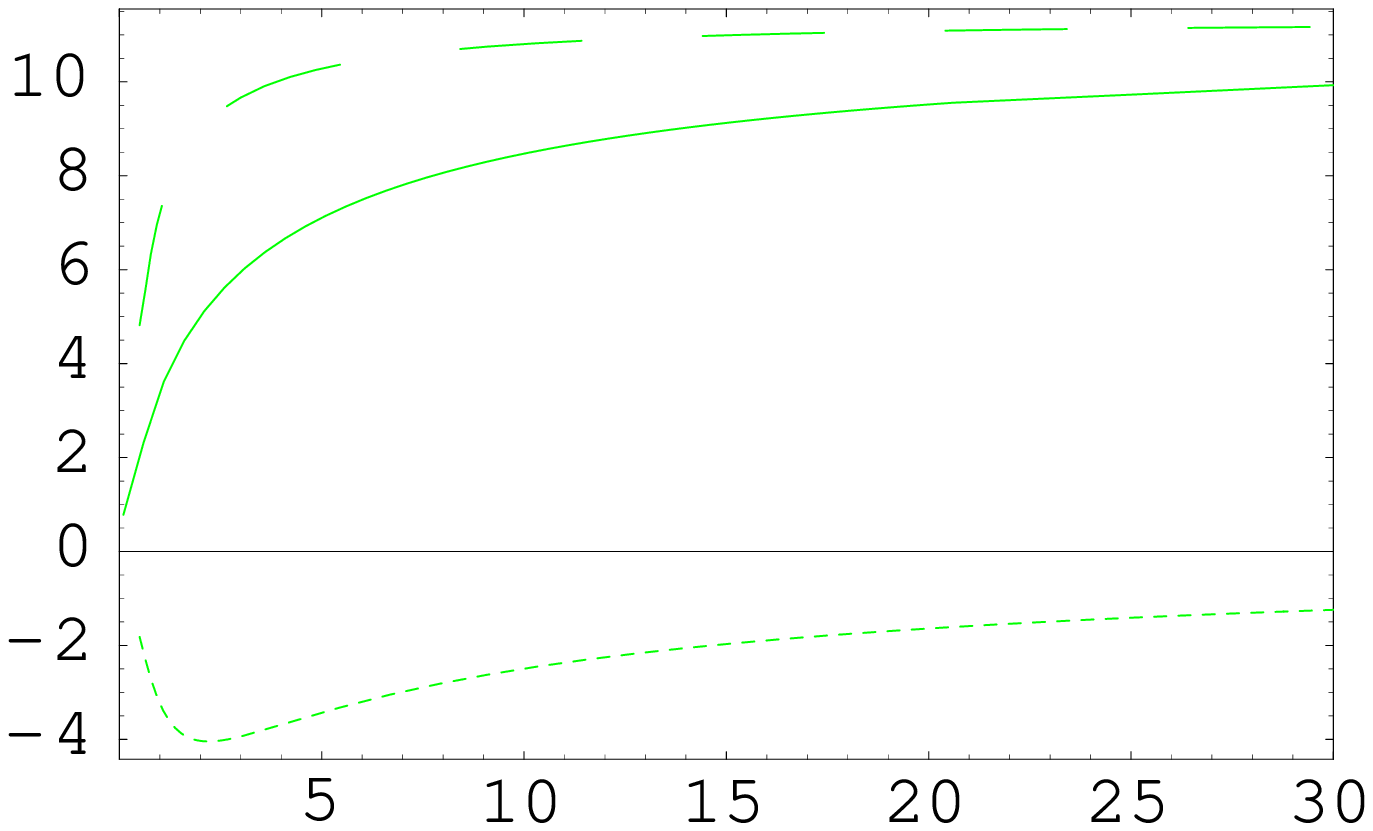,width=50mm,height=50mm}\\

\multicolumn{3}{c}{\rule[-3mm]{0mm}{4mm} $F_T(Q^2)$}\\
$x_B=10^{-2}$&$x_B=5\cdot 10^{-3}$& $x_B=10^{-3}$\\[-10mm]
\epsfig{file=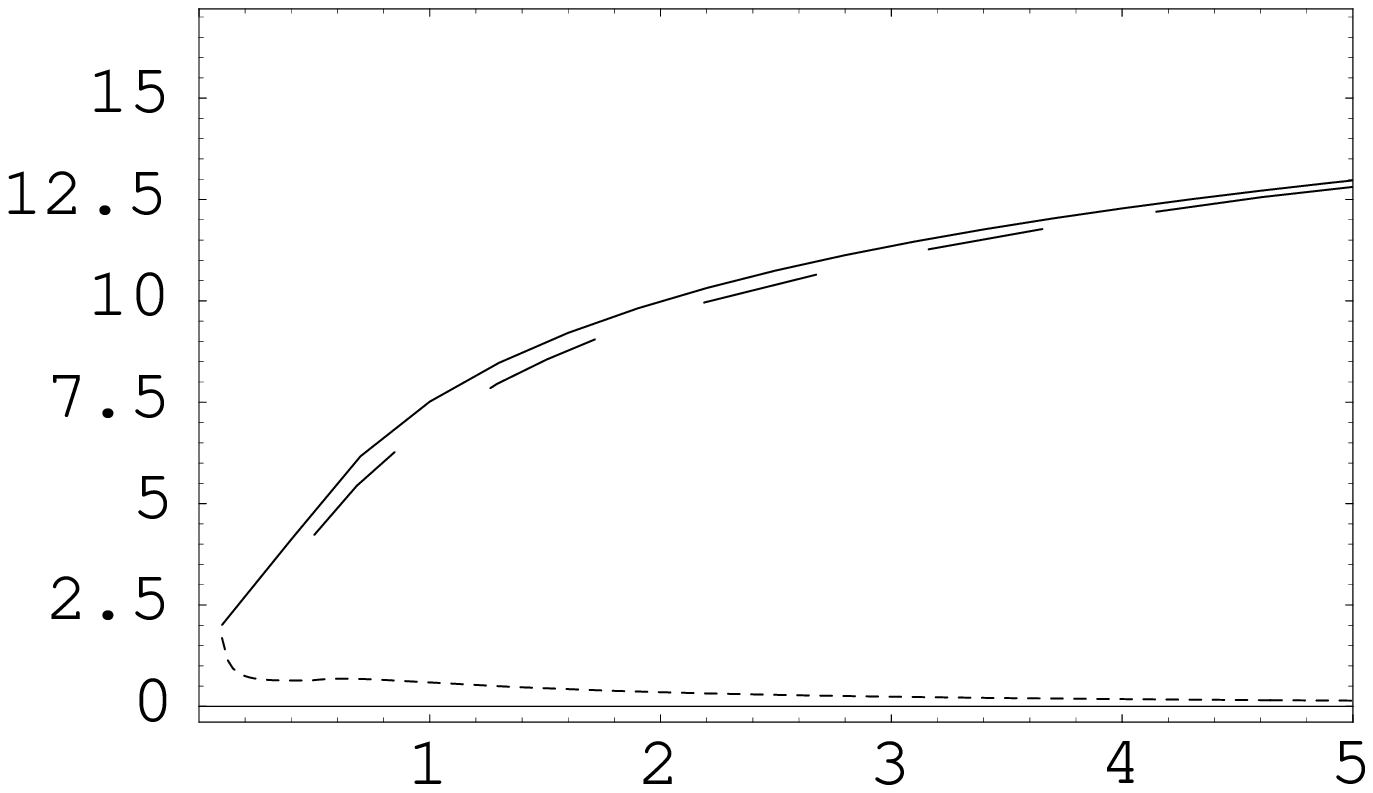,width=50mm,height=50mm} &
\epsfig{file=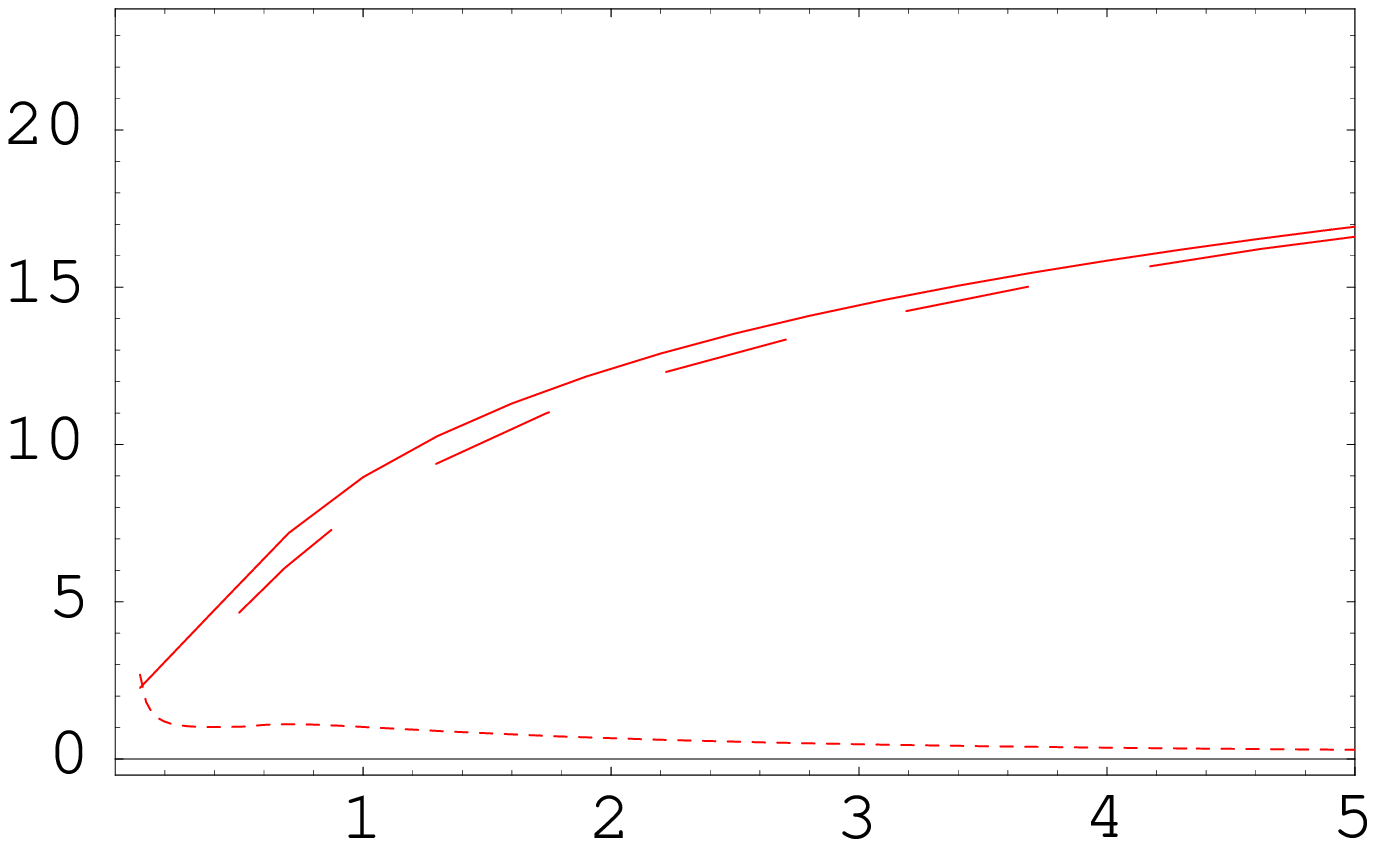,width=50mm,height=50mm} &
\epsfig{file=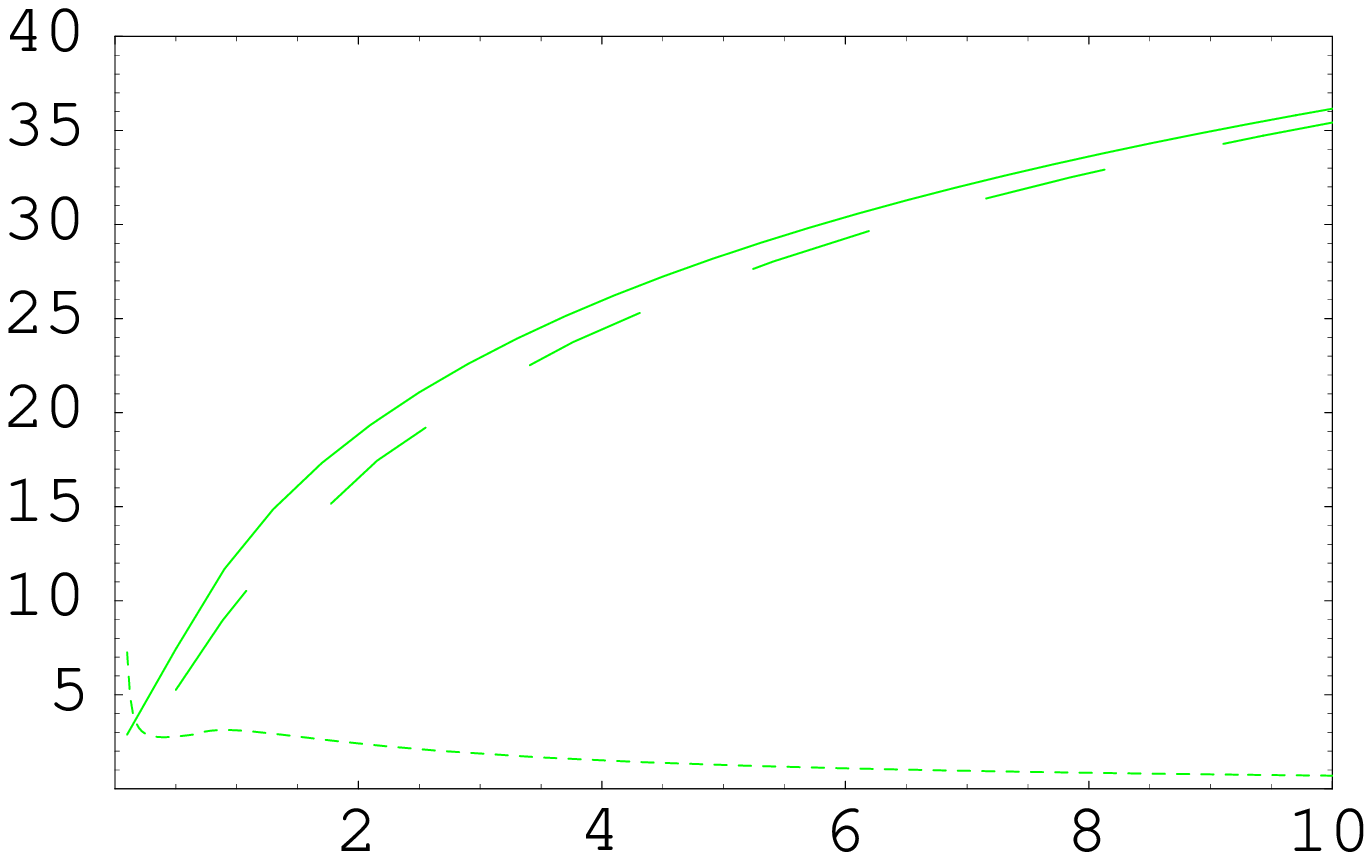,width=50mm,height=50mm}\\

\multicolumn{3}{c}{\rule[-3mm]{0mm}{4mm} $F_L^D(Q^2)$}\\
$x_B=10^{-2}$&$x_B=5\cdot 10^{-3}$& $x_B=10^{-3}$\\[-10mm]
\epsfig{file=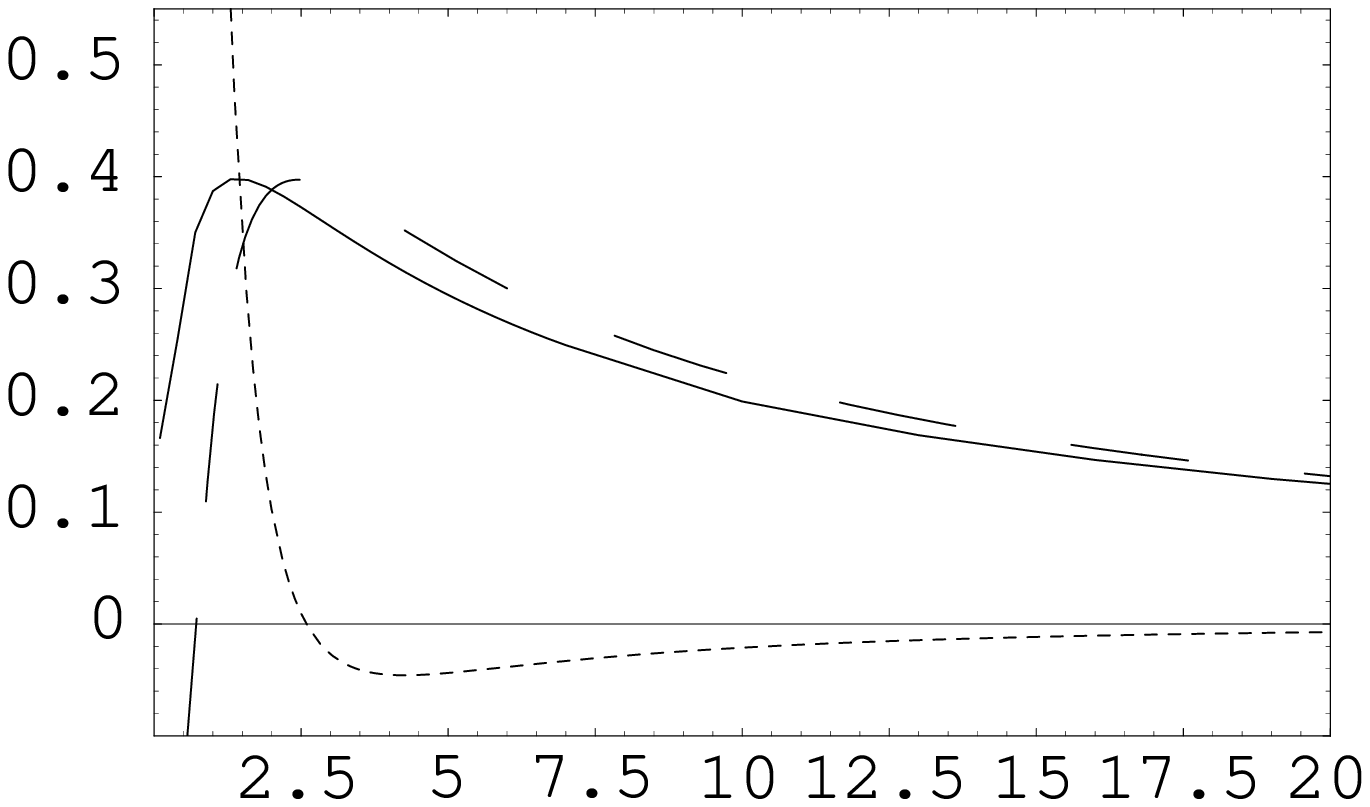,width=50mm,height=50mm} &
\epsfig{file=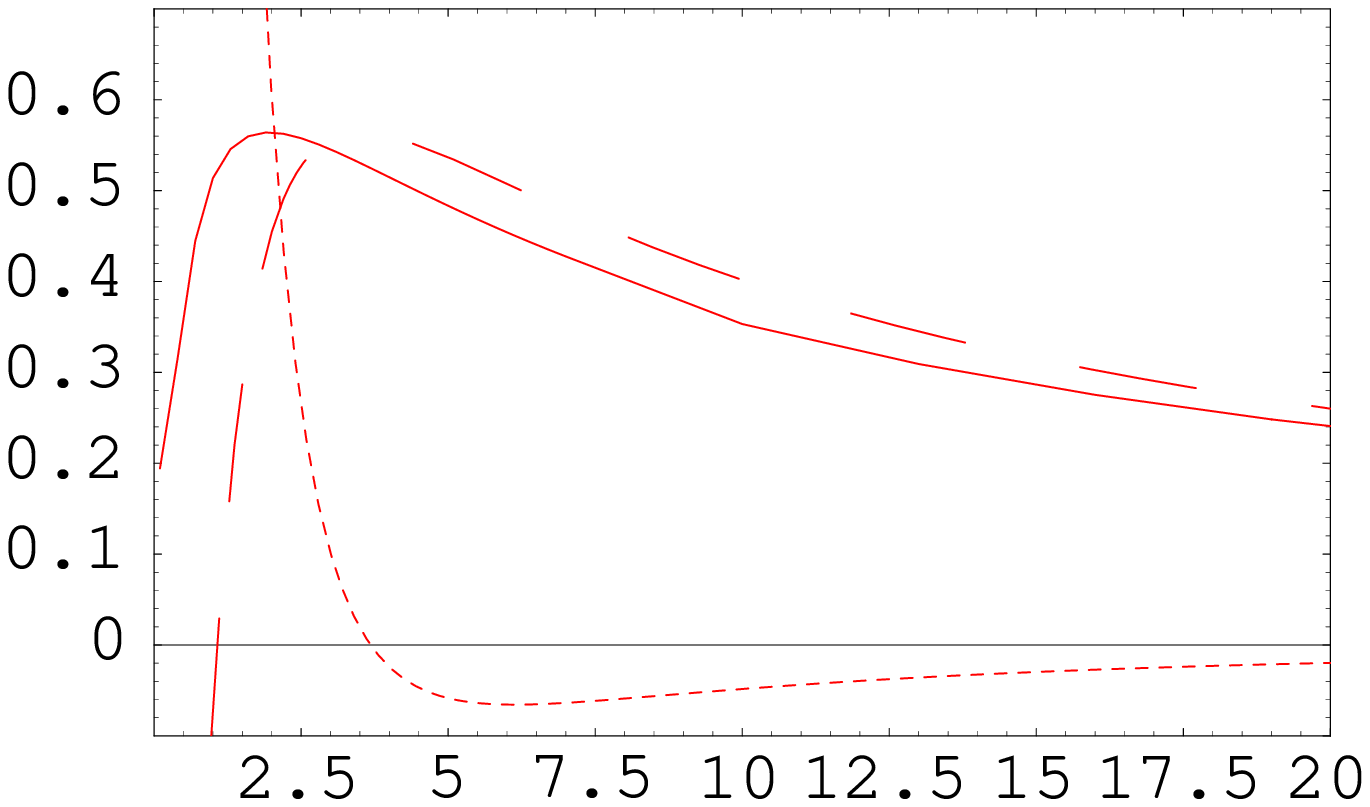,width=50mm,height=50mm} &
\epsfig{file=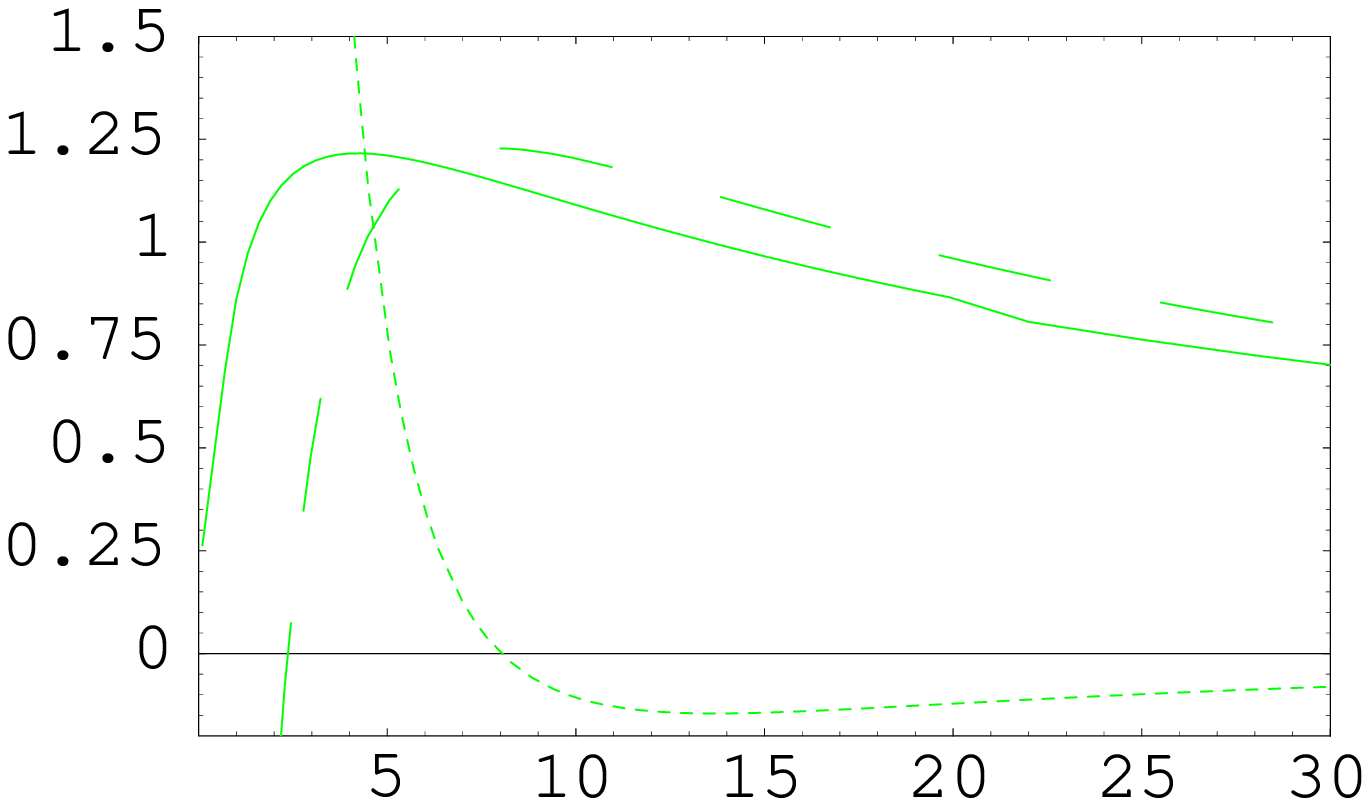,width=50mm,height=50mm}\\

\multicolumn{3}{c}{\rule[-3mm]{0mm}{4mm} $F_T^D(Q^2)$}\\
$x_B=10^{-2}$&$x_B=5\cdot 10^{-3}$& $x_B=10^{-3}$\\[-10mm]
\epsfig{file=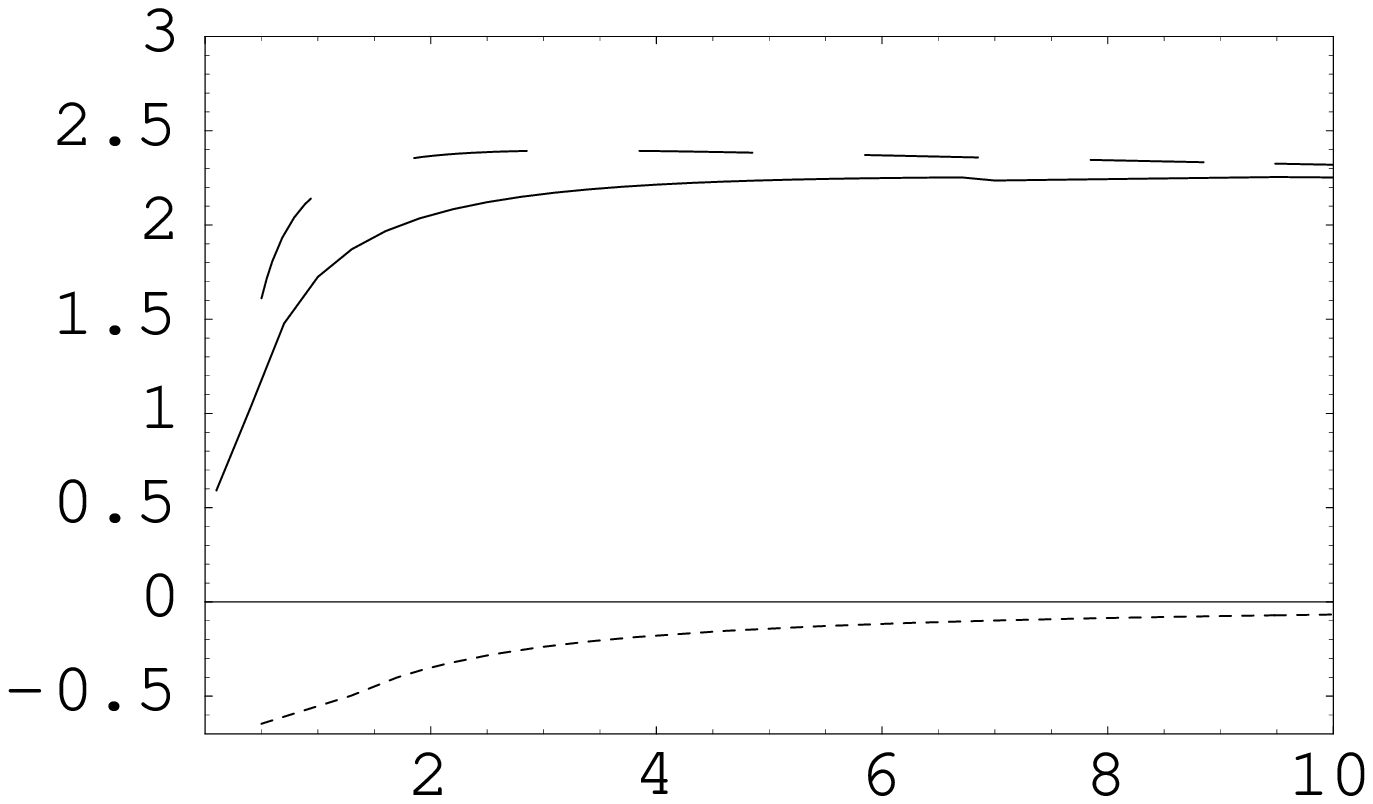,width=50mm,height=50mm} &
\epsfig{file=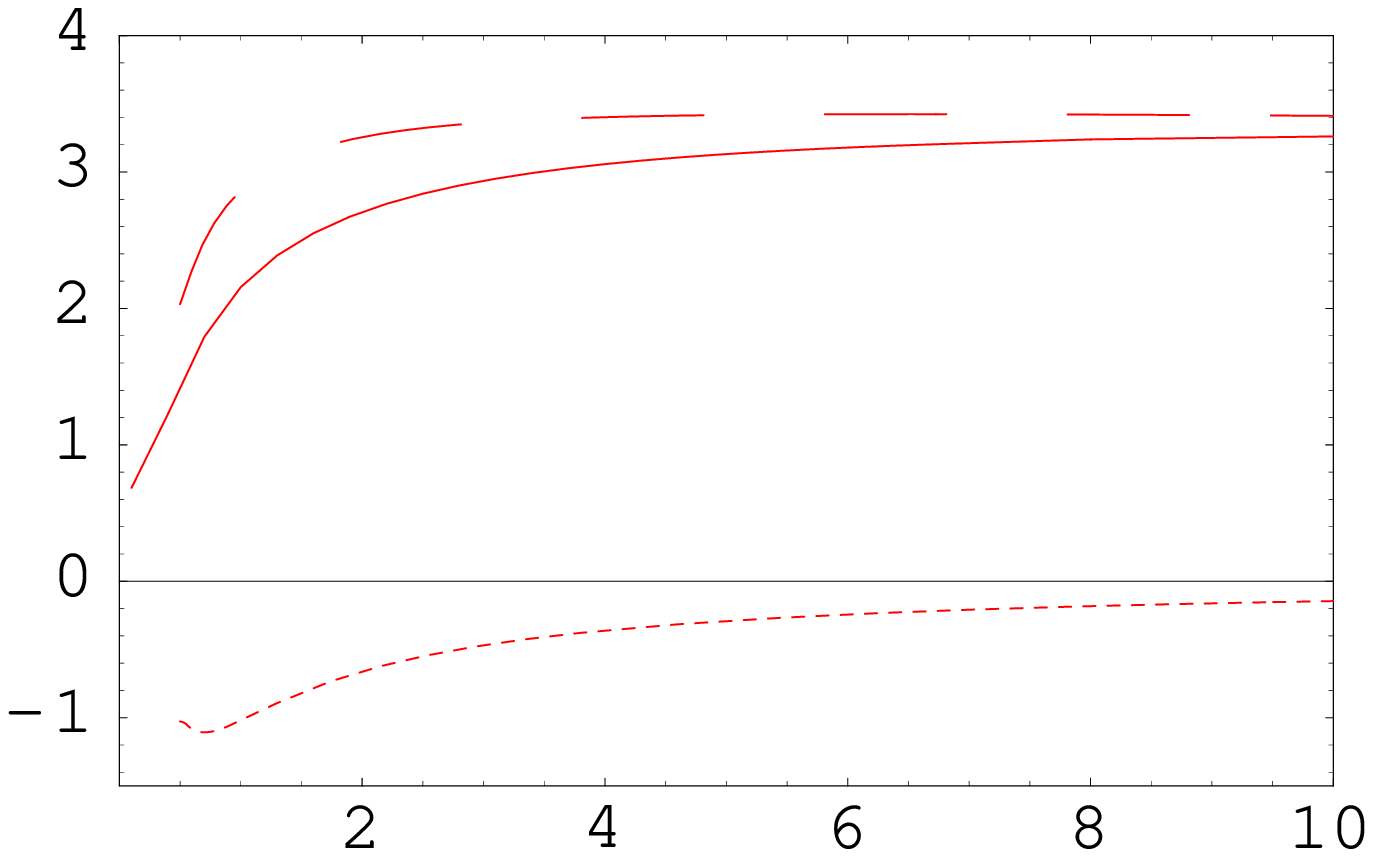,width=50mm,height=50mm} &
\epsfig{file=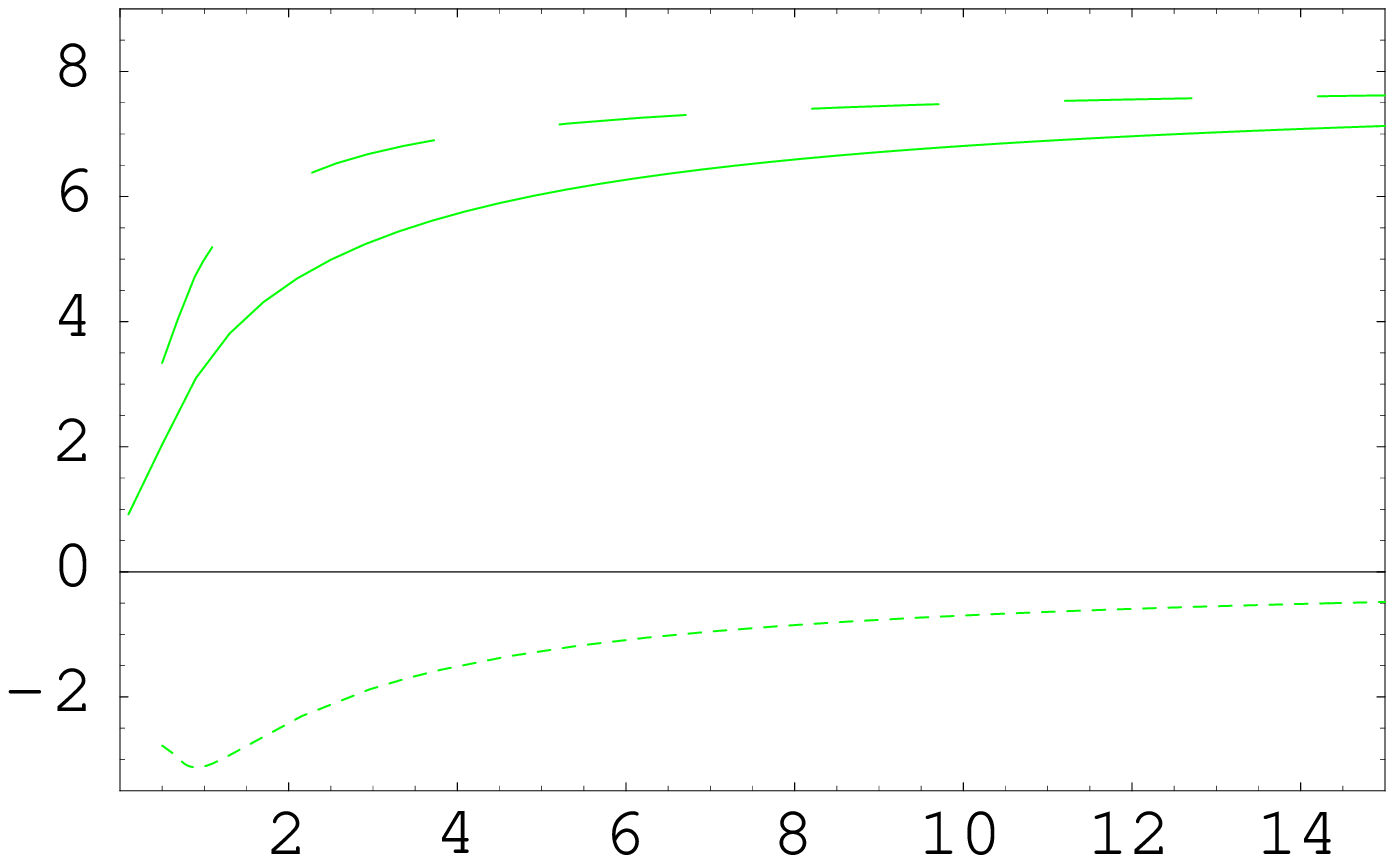,width=50mm,height=50mm}
\end{tabular}
\end{flushleft}
\vspace{-1cm}
\caption{\footnotesize {\it Different twists contributions
to the various structure functions for DIS on nucleus $A$=30:
leading twist (at high $Q^2$) -- dashed line,
next-to-leading -- dotted one, exact structure function --
solid curve.}}
\label{fig2.30}
\end{figure}

\begin{figure}
\begin{flushleft}
\begin{tabular}{ccc}
\multicolumn{3}{c}{\rule[-3mm]{0mm}{4mm} $ F_L(Q^2)$}\\
$x_B=10^{-2}$&$x_B=5\cdot 10^{-3}$& $x_B=10^{-3}$\\[-10mm]
\epsfig{file=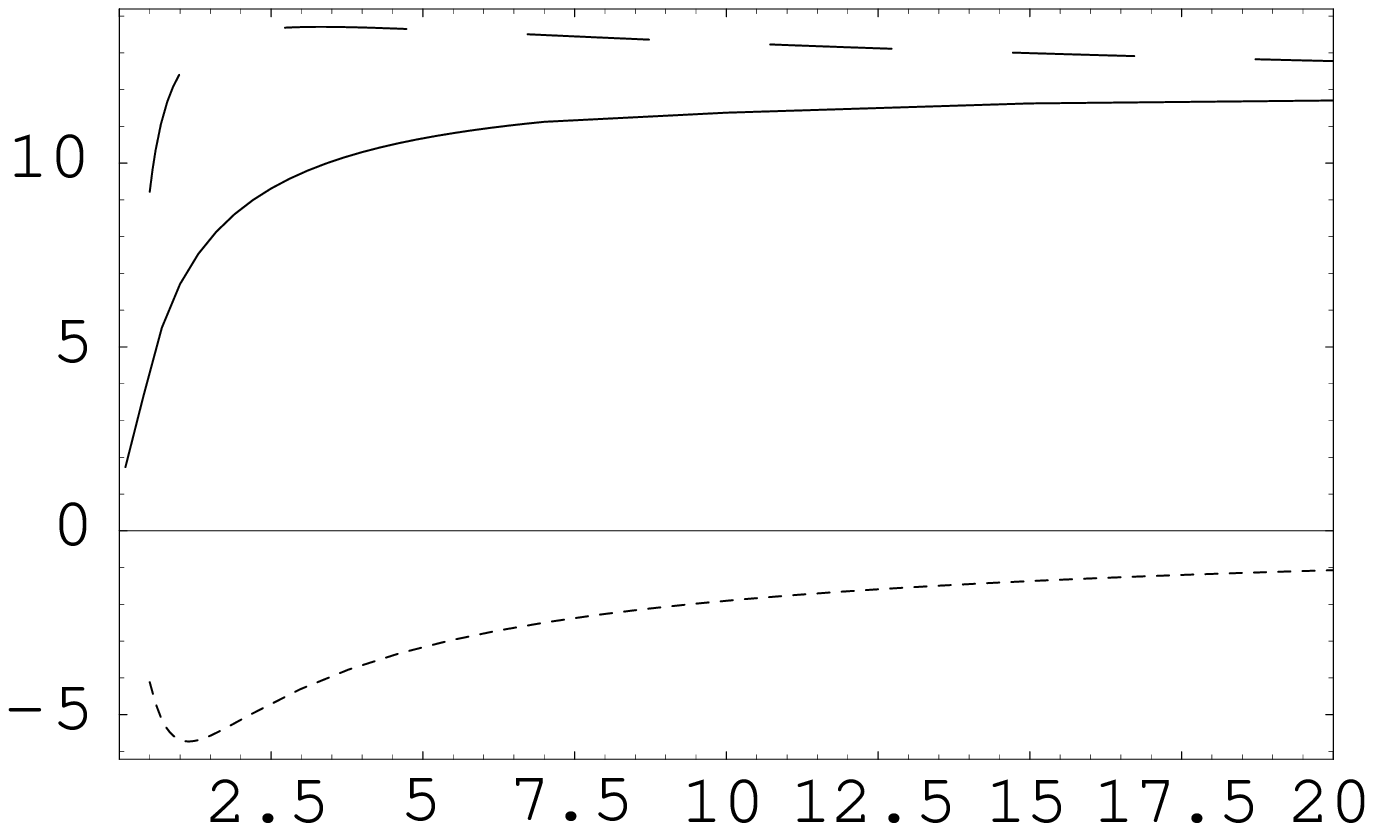,width=50mm,height=50mm} &
\epsfig{file=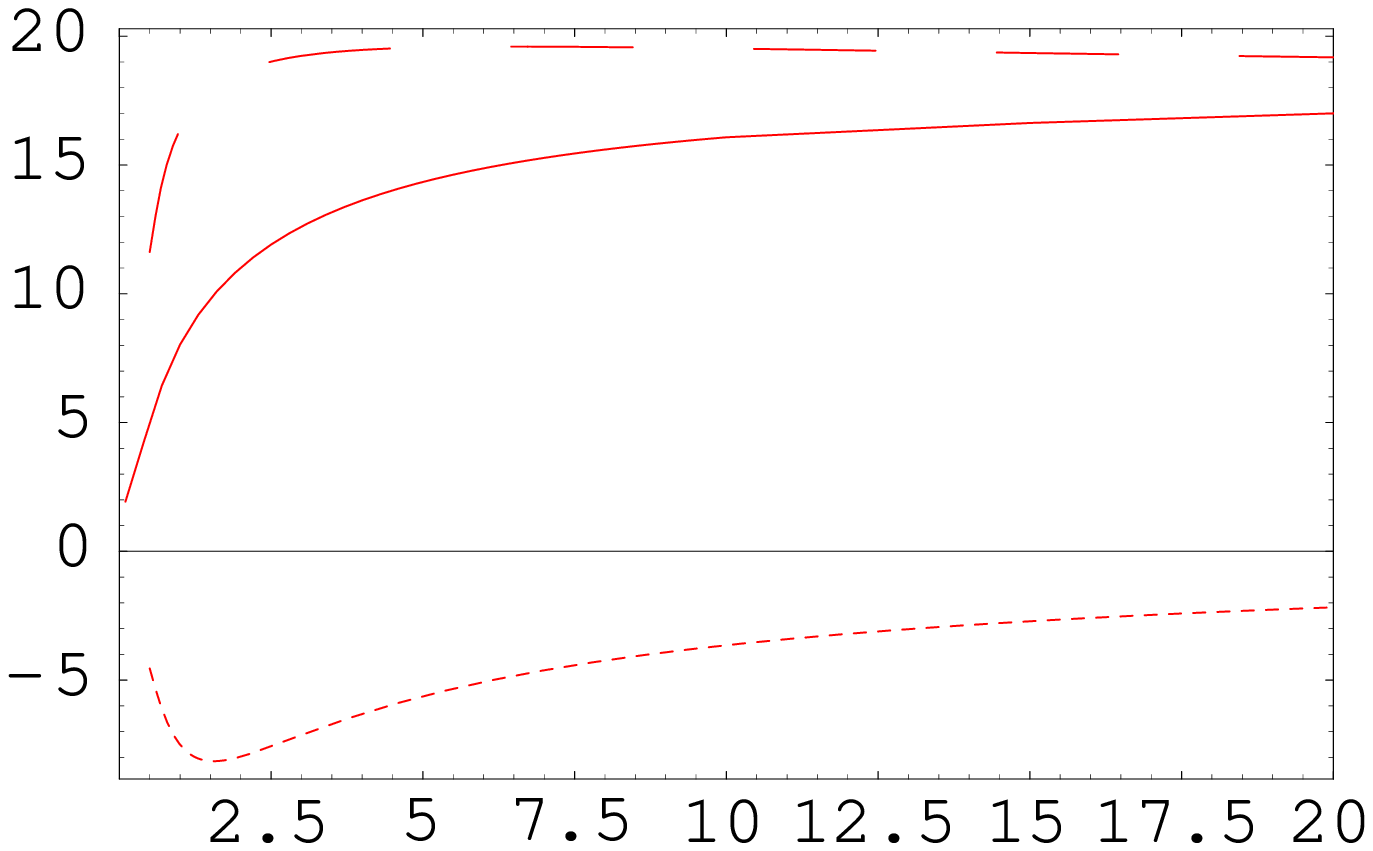,width=50mm,height=50mm}&
\epsfig{file=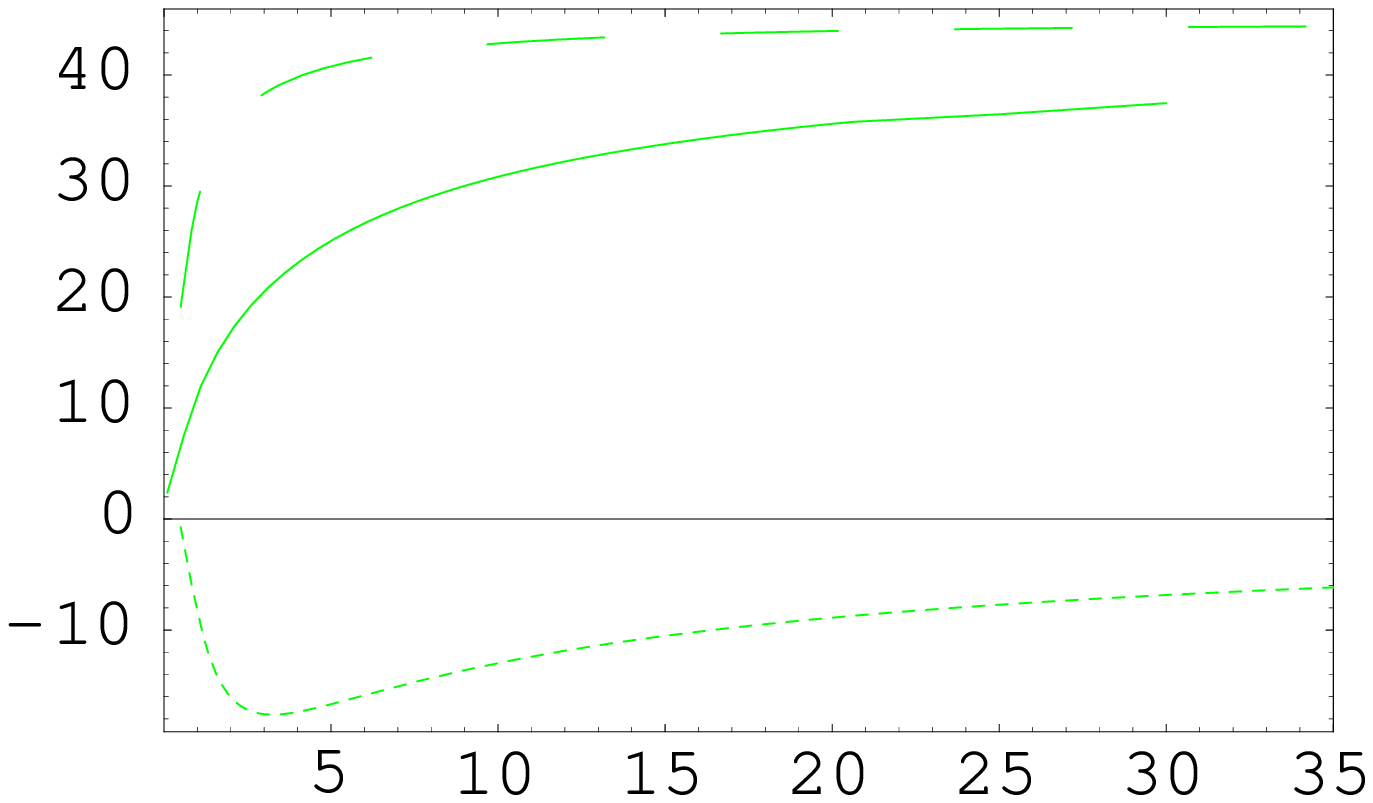,width=50mm,height=50mm}\\

\multicolumn{3}{c}{\rule[-3mm]{0mm}{4mm} $ F_T(Q^2)$}\\
$x_B=10^{-2}$&$x_B=5\cdot 10^{-3}$& $x_B=10^{-3}$\\[-10mm]
\epsfig{file=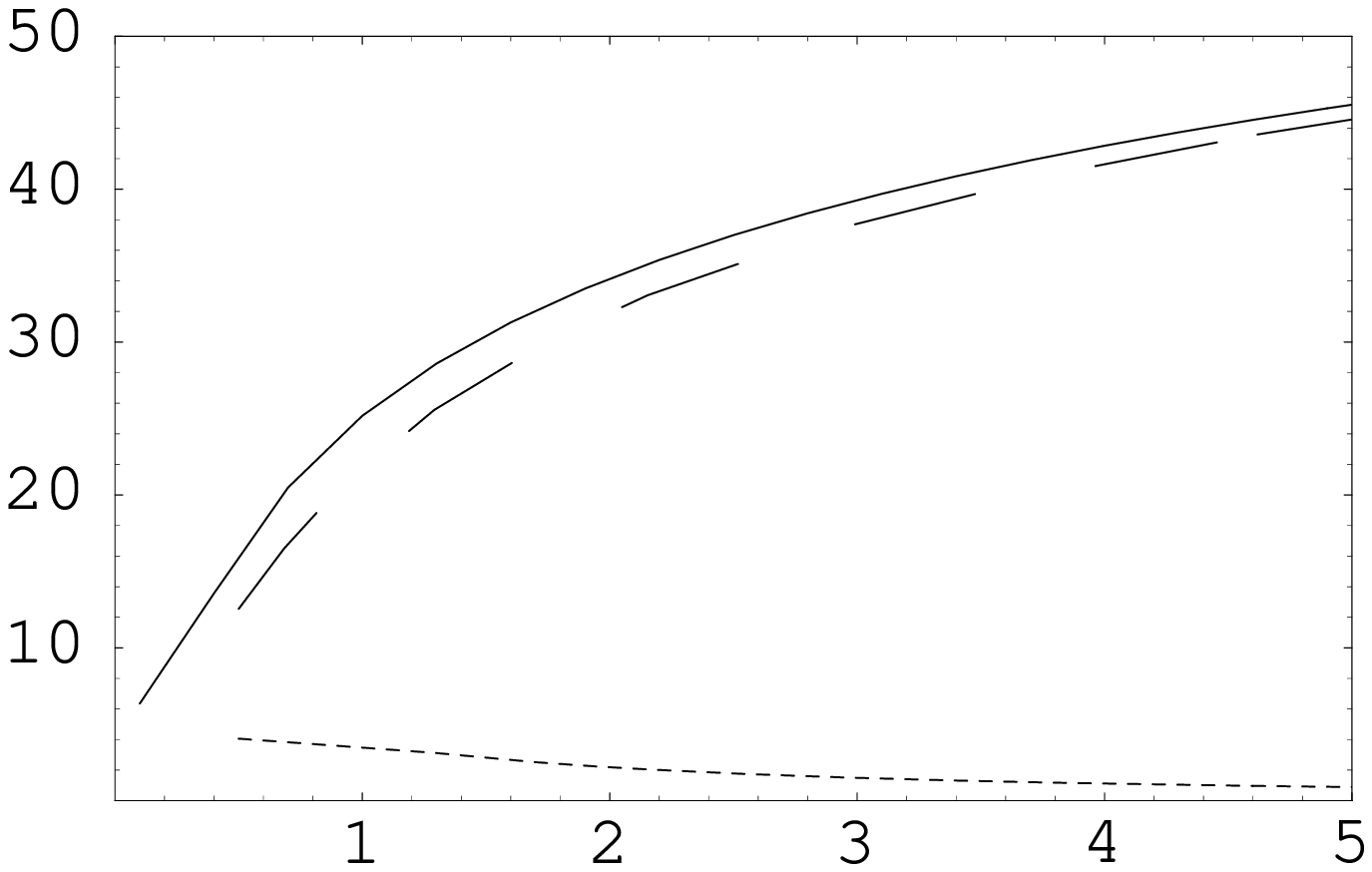,width=50mm,height=50mm} &
\epsfig{file=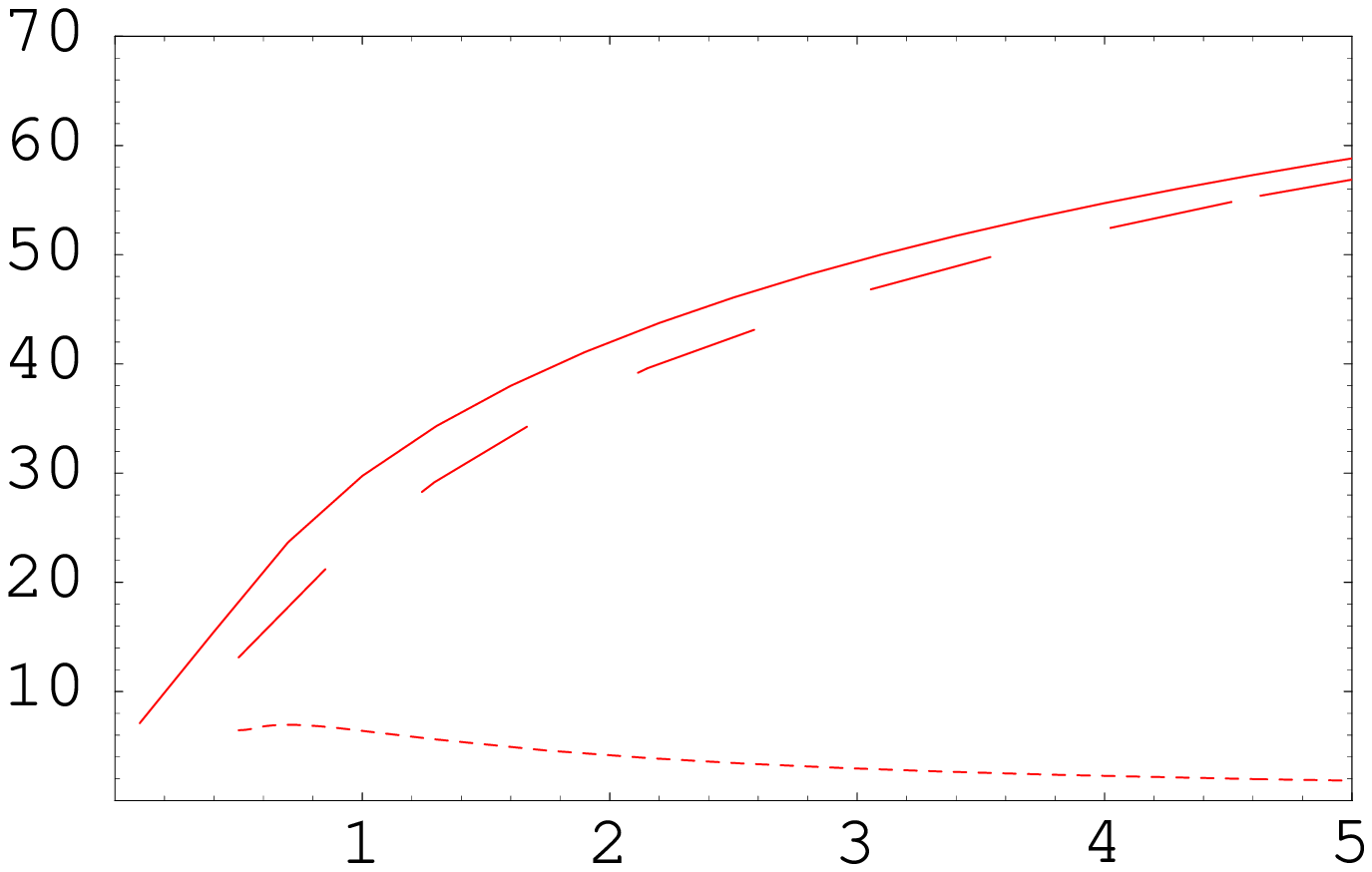,width=50mm,height=50mm} &
\epsfig{file=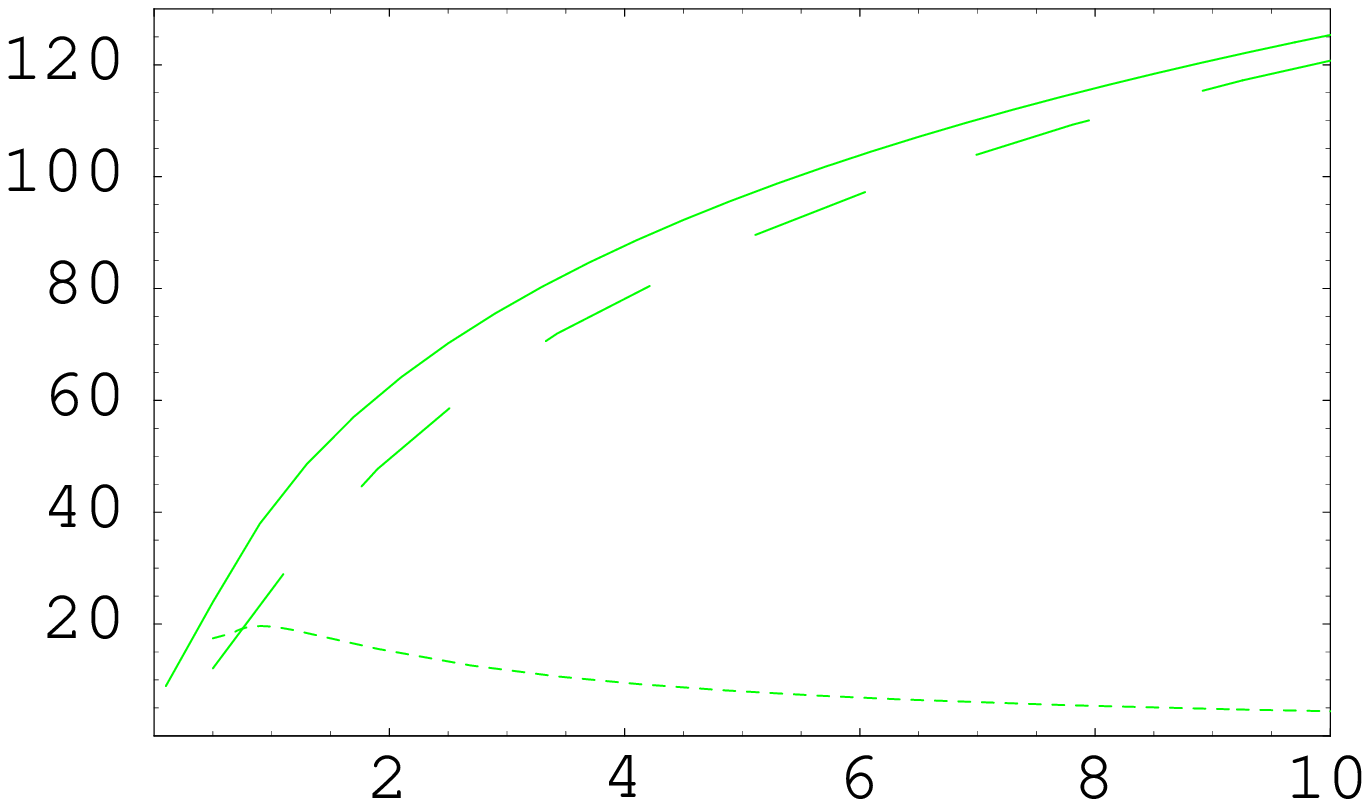,width=50mm,height=50mm}\\

\multicolumn{3}{c}{\rule[-3mm]{0mm}{4mm} $ F_L^D(Q^2)$}\\
$x_B=10^{-2}$&$x_B=5\cdot 10^{-3}$& $x_B=10^{-3}$\\[-10mm]
\epsfig{file=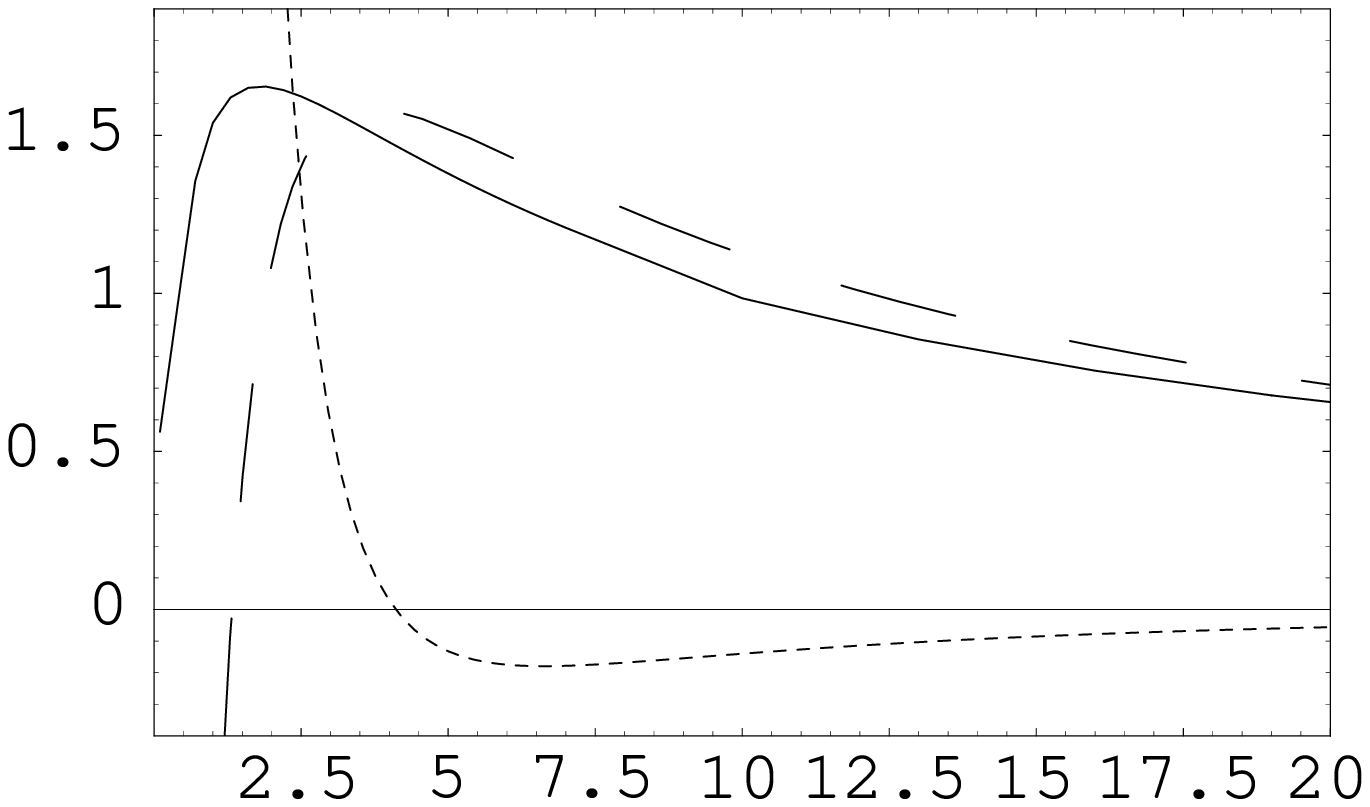,width=50mm,height=50mm} &
\epsfig{file=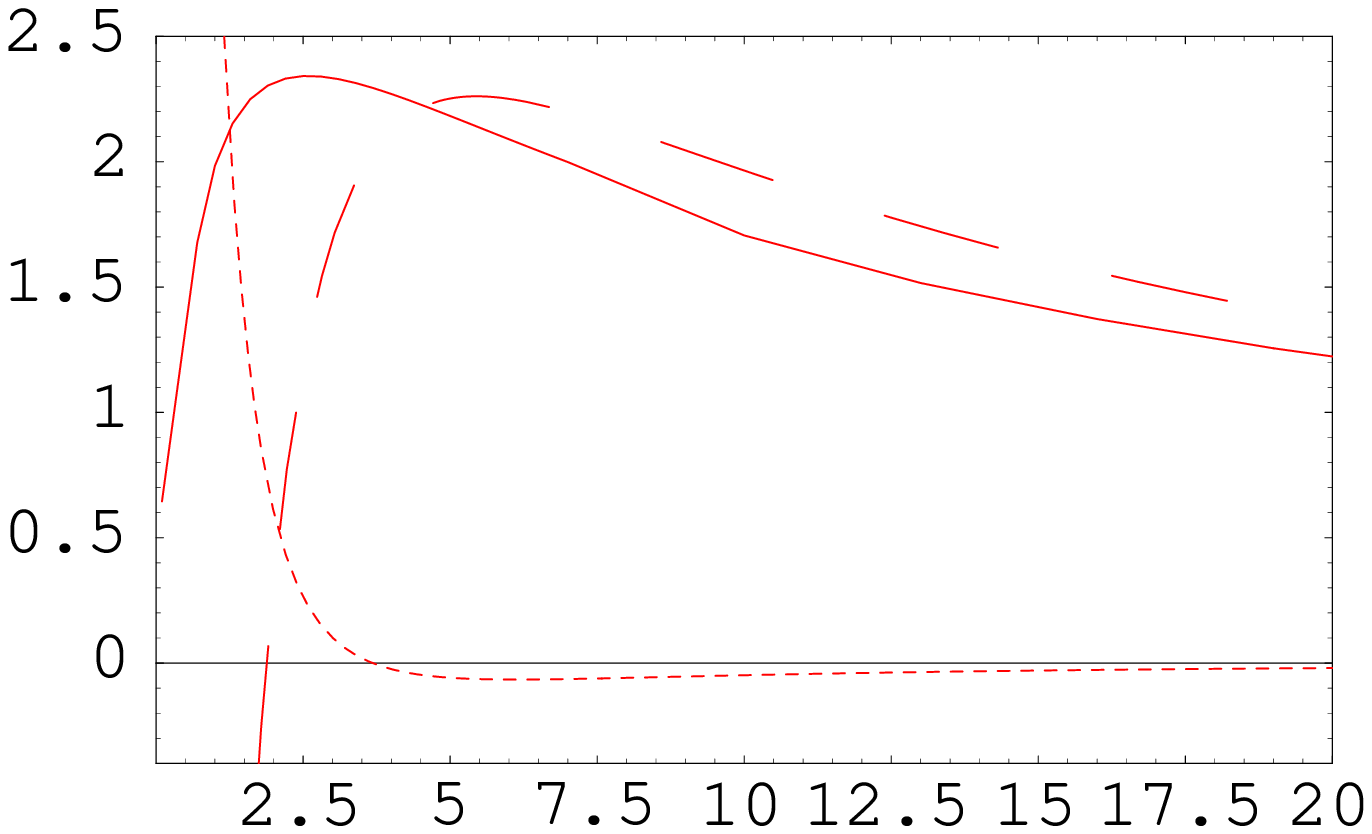,width=50mm,height=50mm} &
\epsfig{file=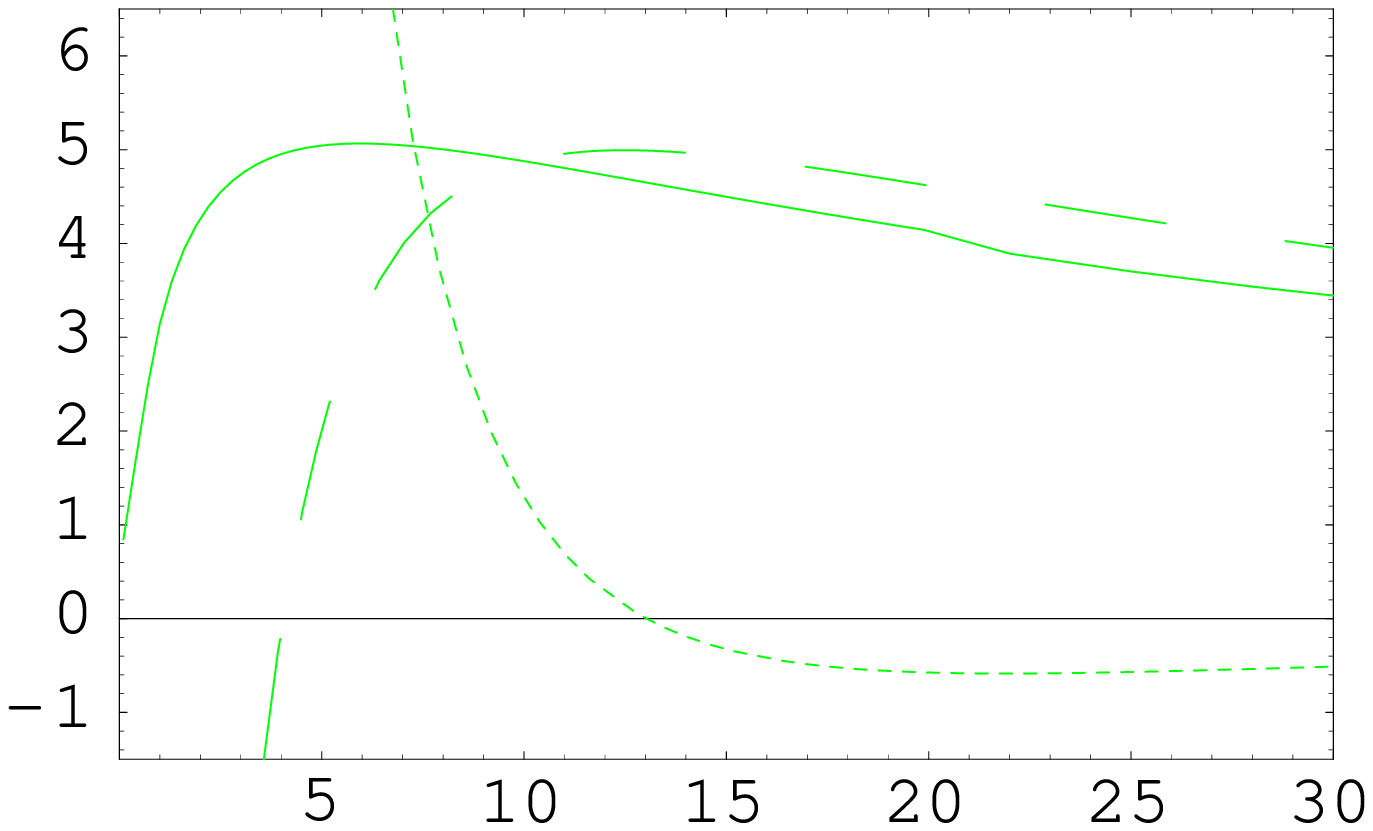,width=50mm,height=50mm}\\

\multicolumn{3}{c}{\rule[-3mm]{0mm}{4mm} $ F_L^D(Q^2)$}\\
$x_B=10^{-2}$&$x_B=5\cdot 10^{-3}$& $x_B=10^{-3}$\\[-10mm]
\epsfig{file=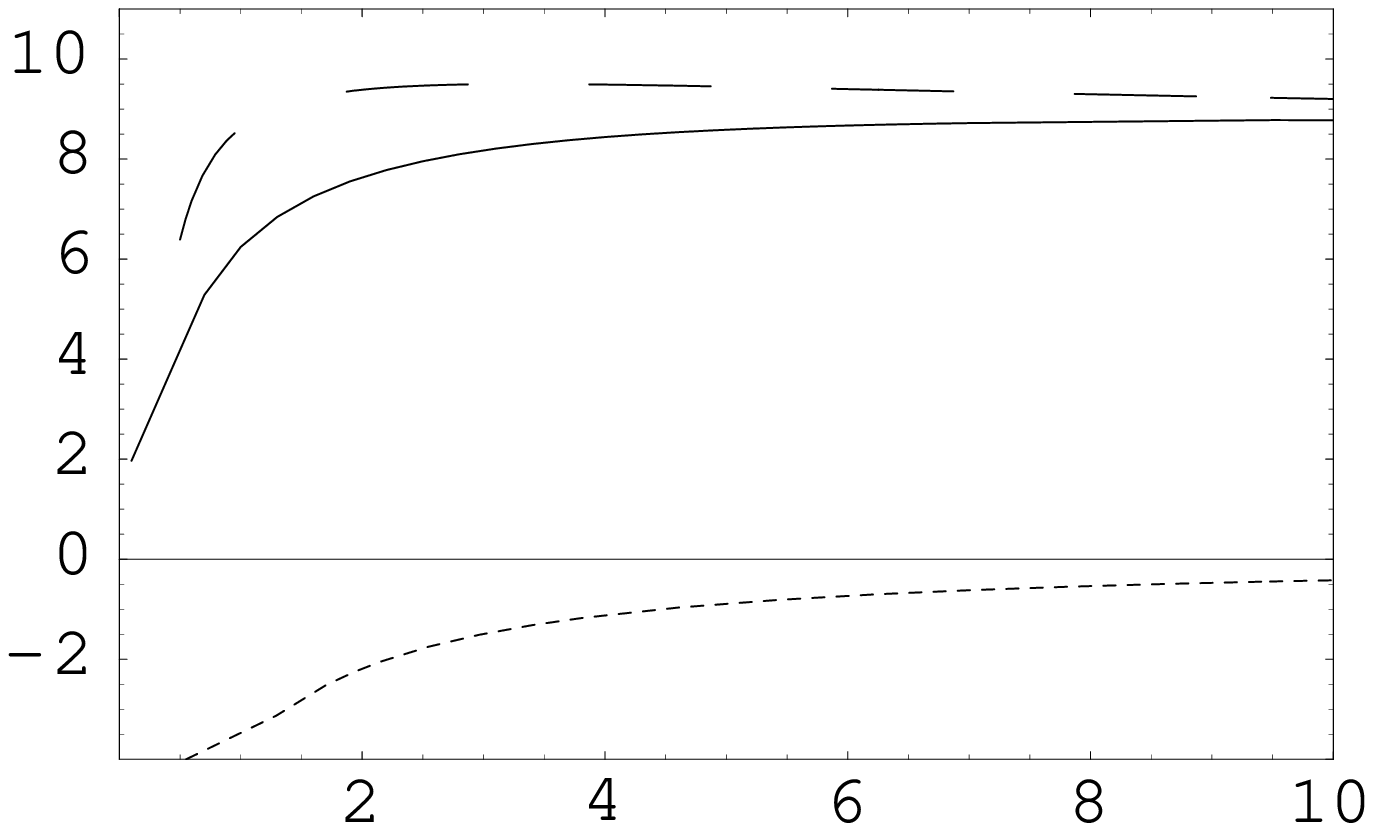,width=50mm,height=50mm} &
\epsfig{file=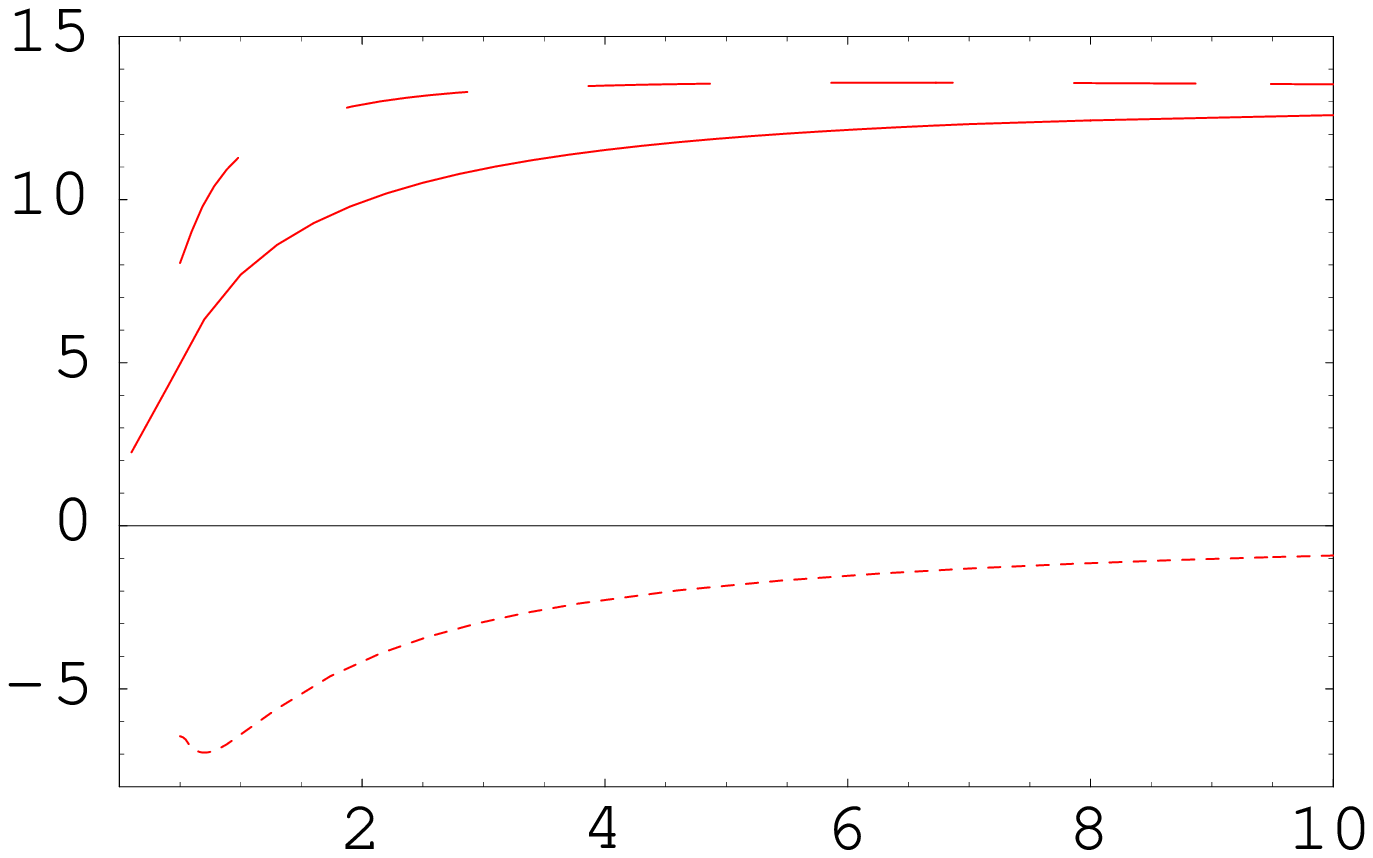,width=50mm,height=50mm} &
\epsfig{file=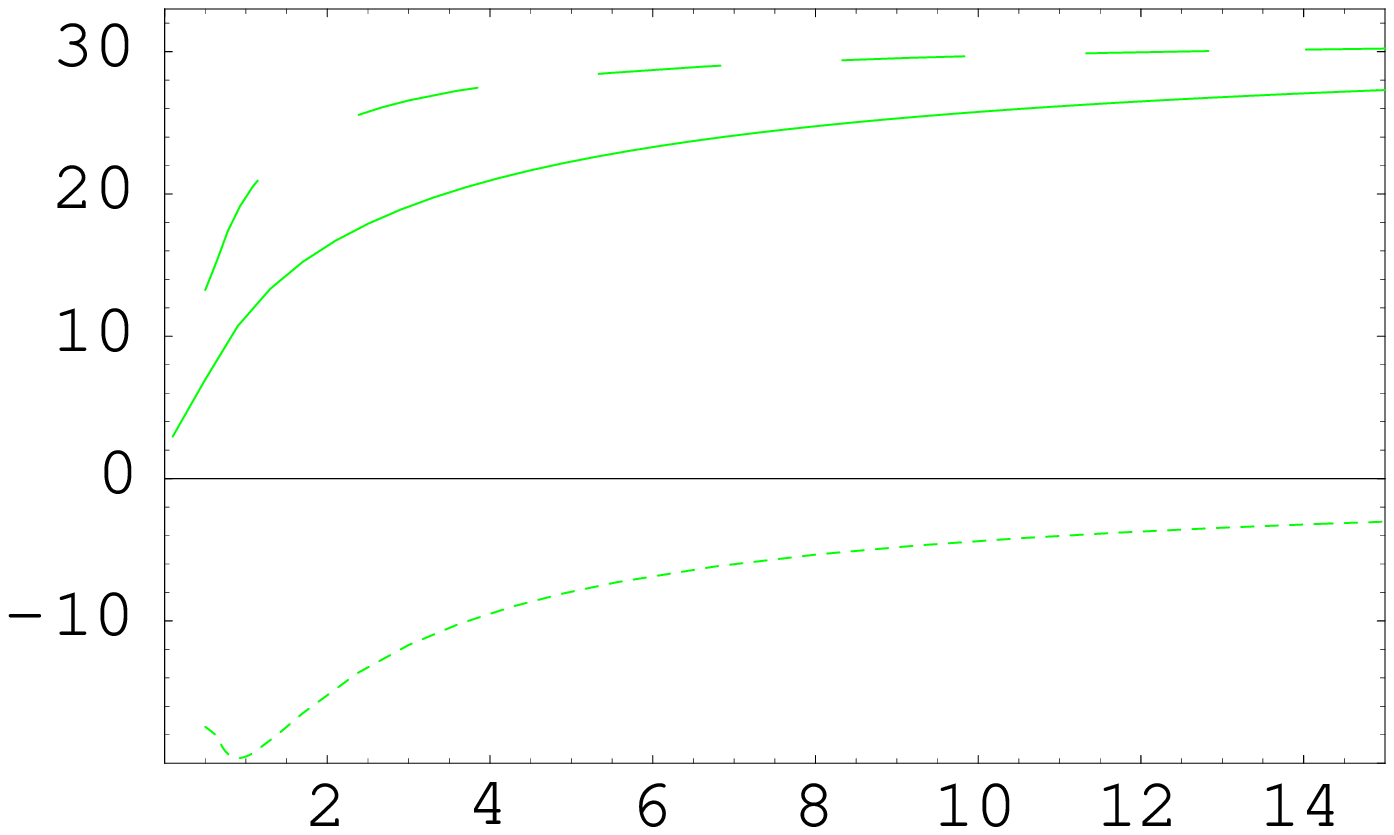,width=50mm,height=50mm}
\end{tabular}
\end{flushleft}
\vspace{-1cm}
\caption{\footnotesize {\it Different twists contributions
to the various structure functions for DIS on nucleus $A$=119:
leading twist (at high $Q^2$) -- dashed line,
next-to-leading -- dotted one, exact structure function --
solid curve.}}
\label{fig2.119}
\end{figure}

\begin{figure}
\begin{flushleft}
\begin{tabular}{ccc}
\multicolumn{3}{c}{\rule[-3mm]{0mm}{4mm} $ F_L(Q^2)$}\\
$x_B=10^{-2}$&$x_B=5\cdot 10^{-3}$& $x_B=10^{-3}$\\[-10mm]
\epsfig{file=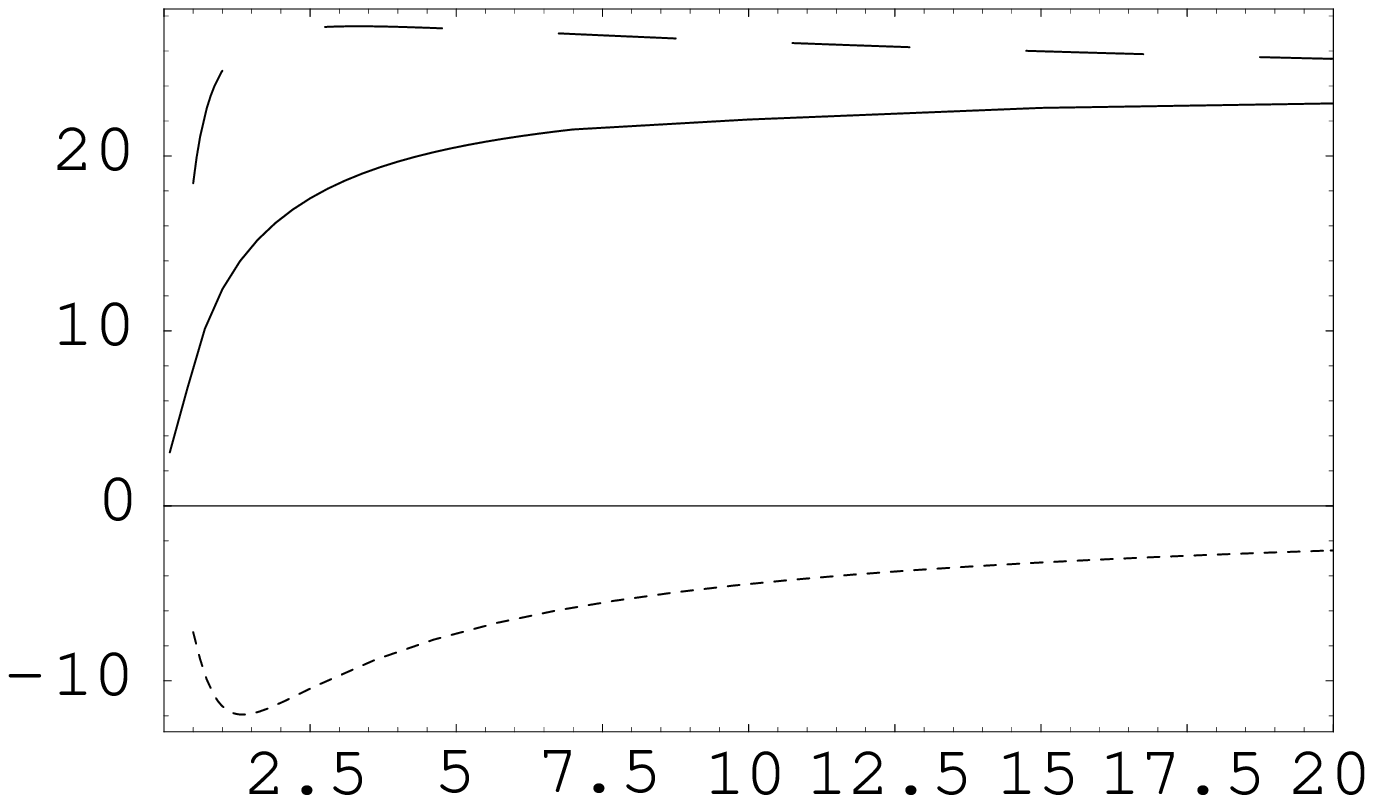,width=50mm,height=50mm} &
\epsfig{file=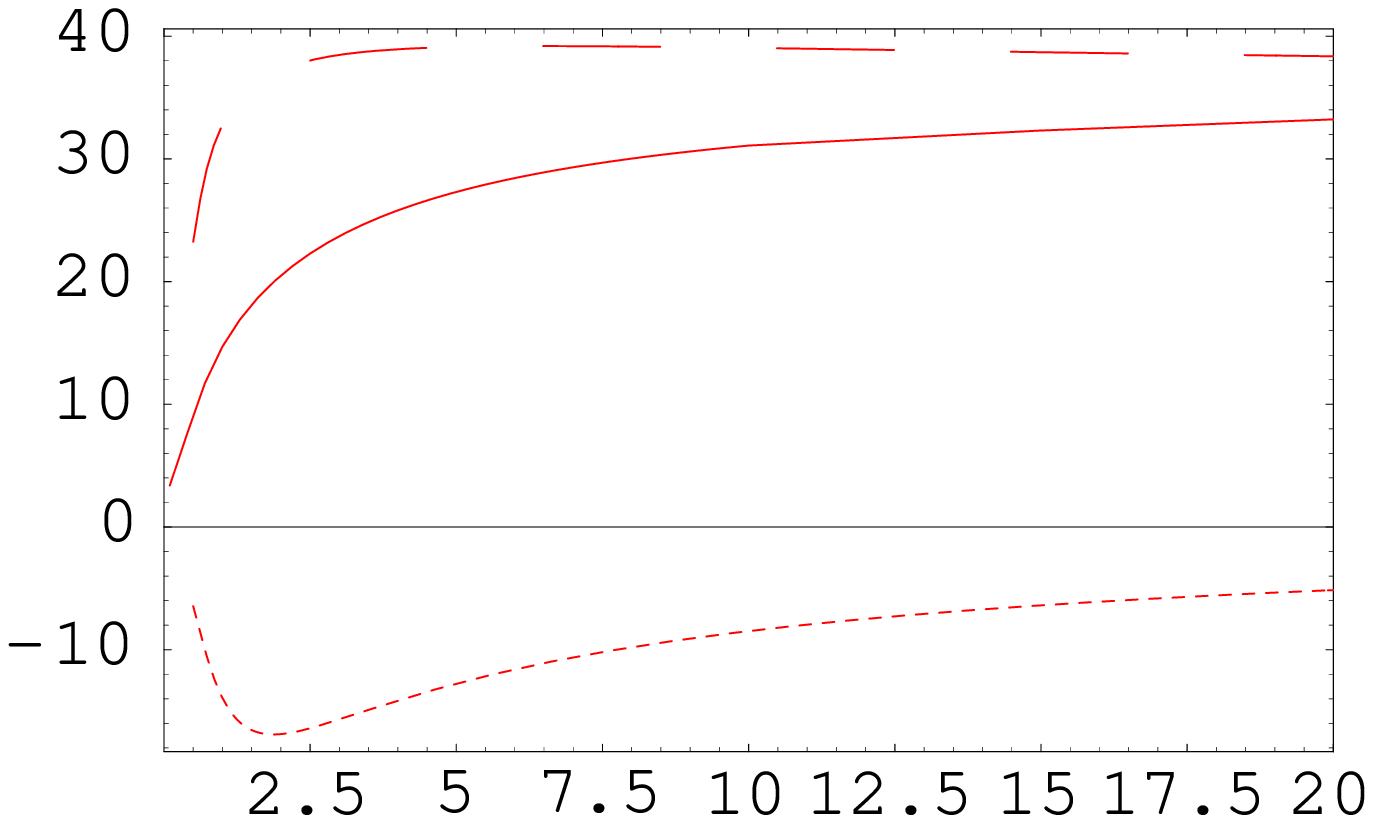,width=50mm,height=50mm}&
\epsfig{file=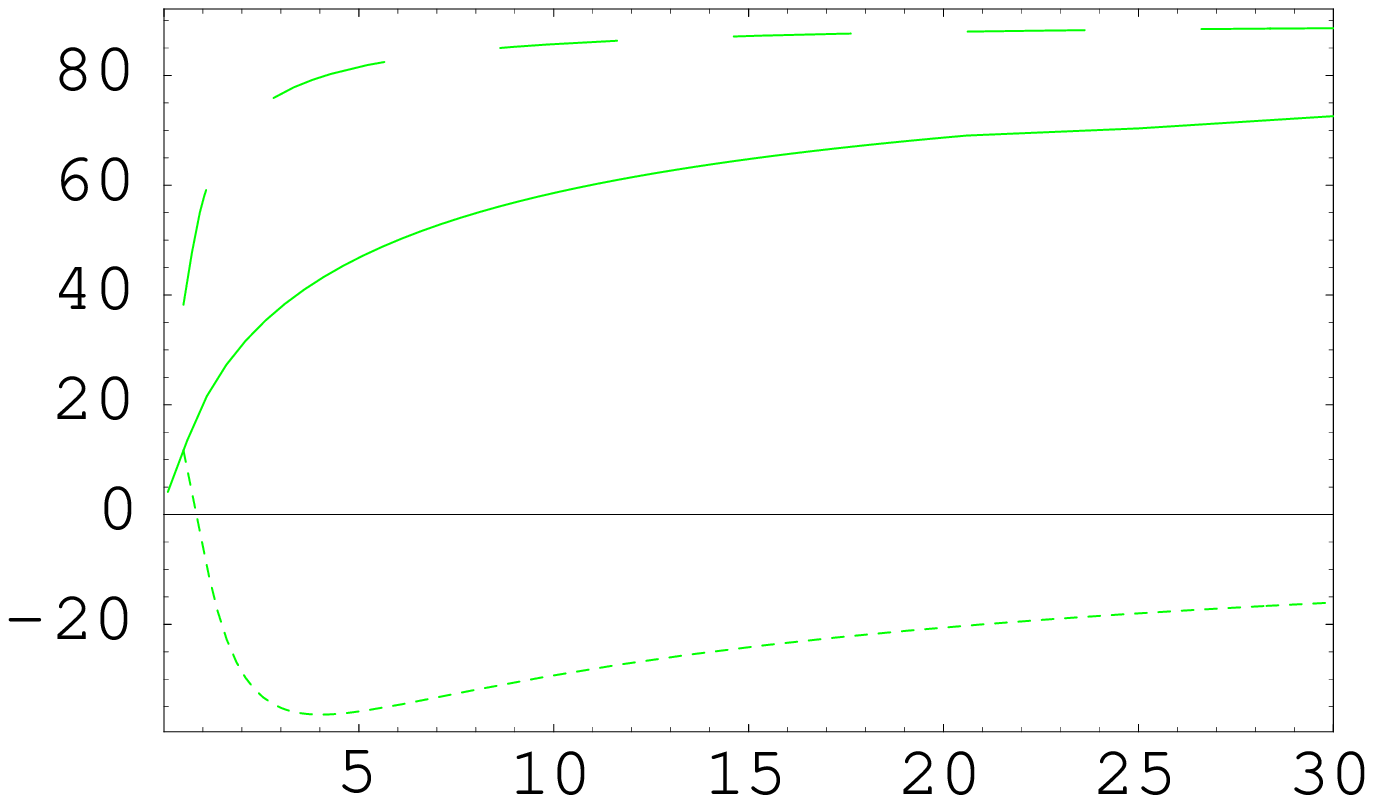,width=50mm,height=50mm}\\

\multicolumn{3}{c}{\rule[-3mm]{0mm}{4mm} $ F_T(Q^2)$}\\
$x_B=10^{-2}$&$x_B=5\cdot 10^{-3}$& $x_B=10^{-3}$\\[-10mm]
\epsfig{file=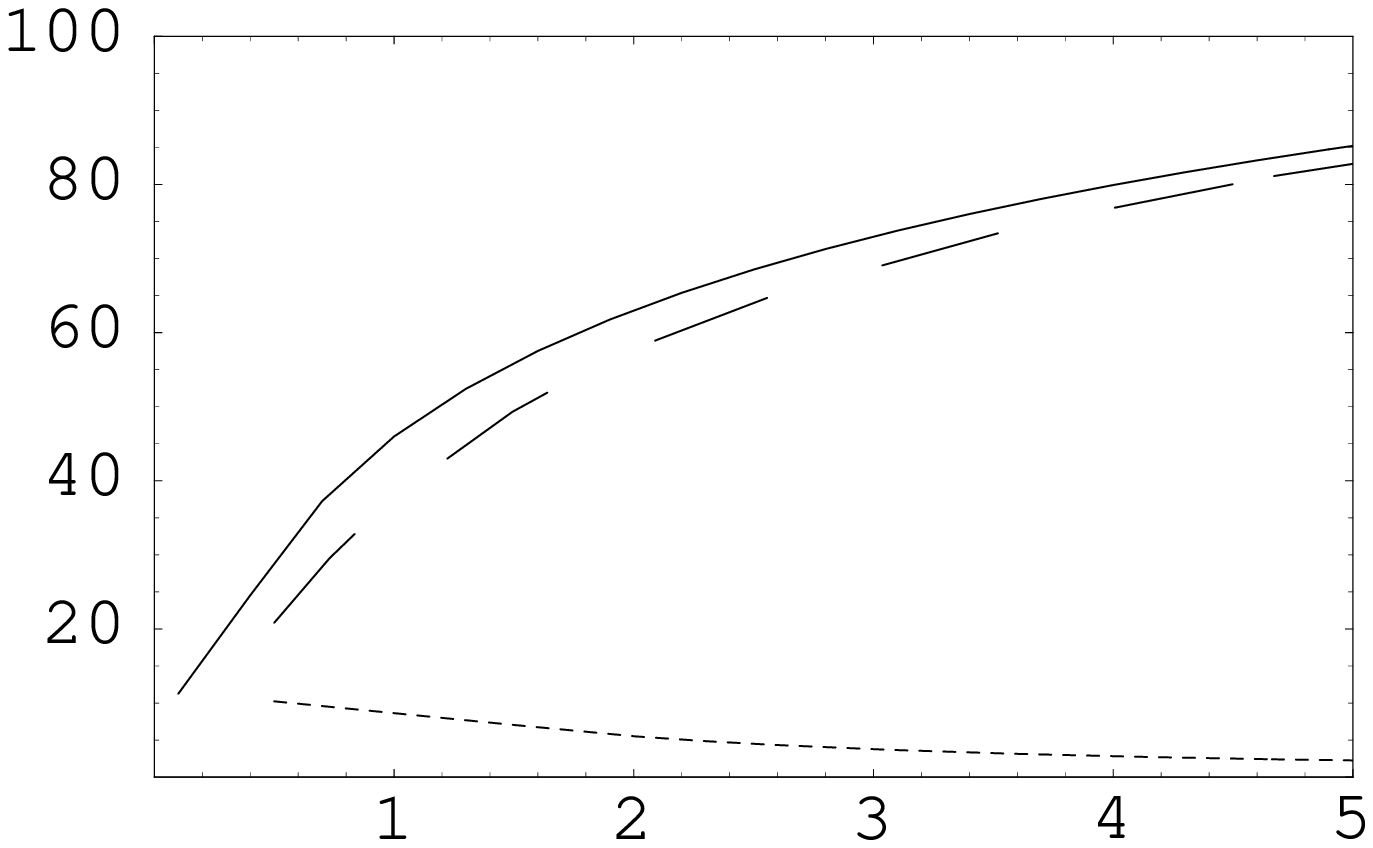,width=50mm,height=50mm} &
\epsfig{file=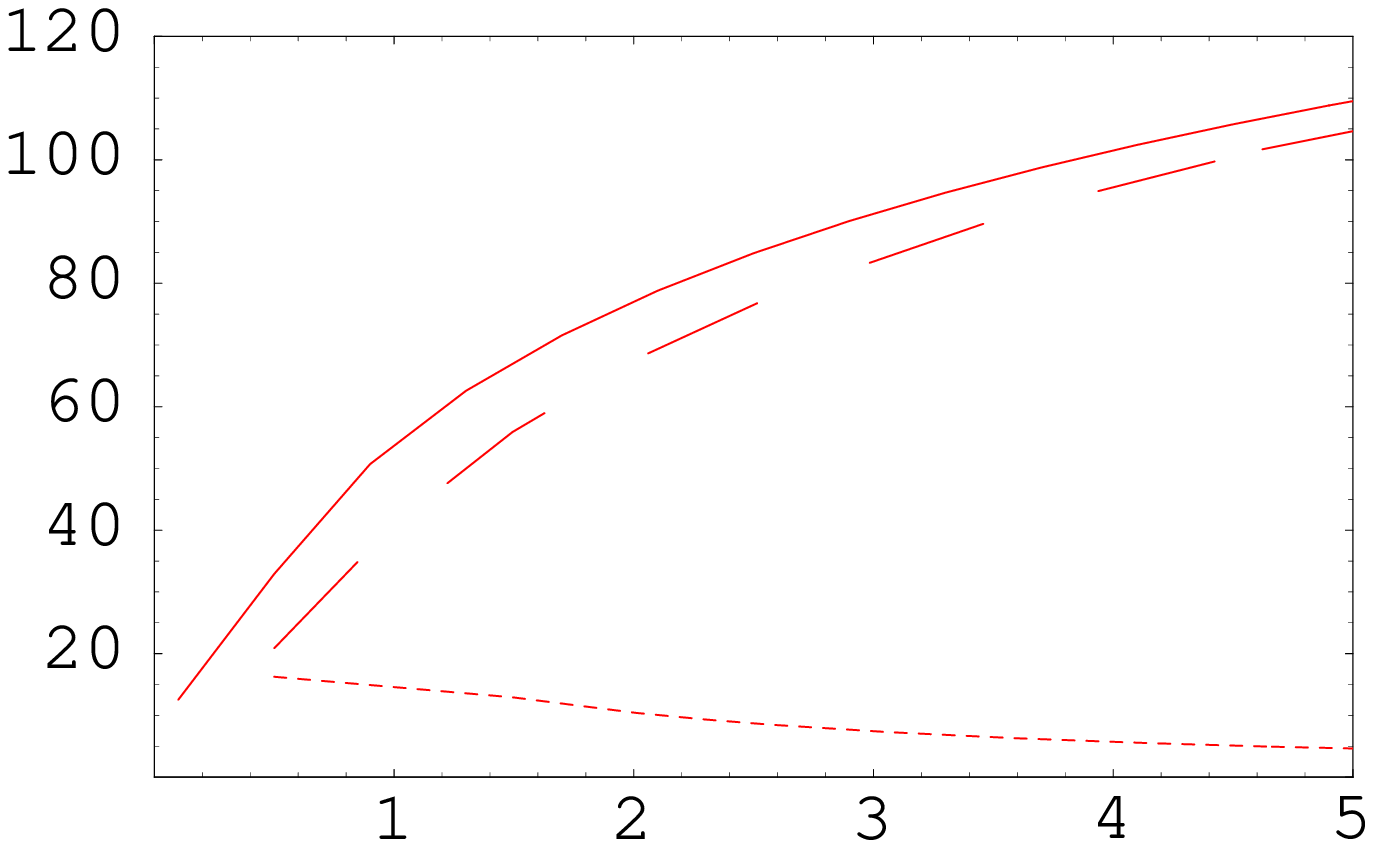,width=50mm,height=50mm} &
\epsfig{file=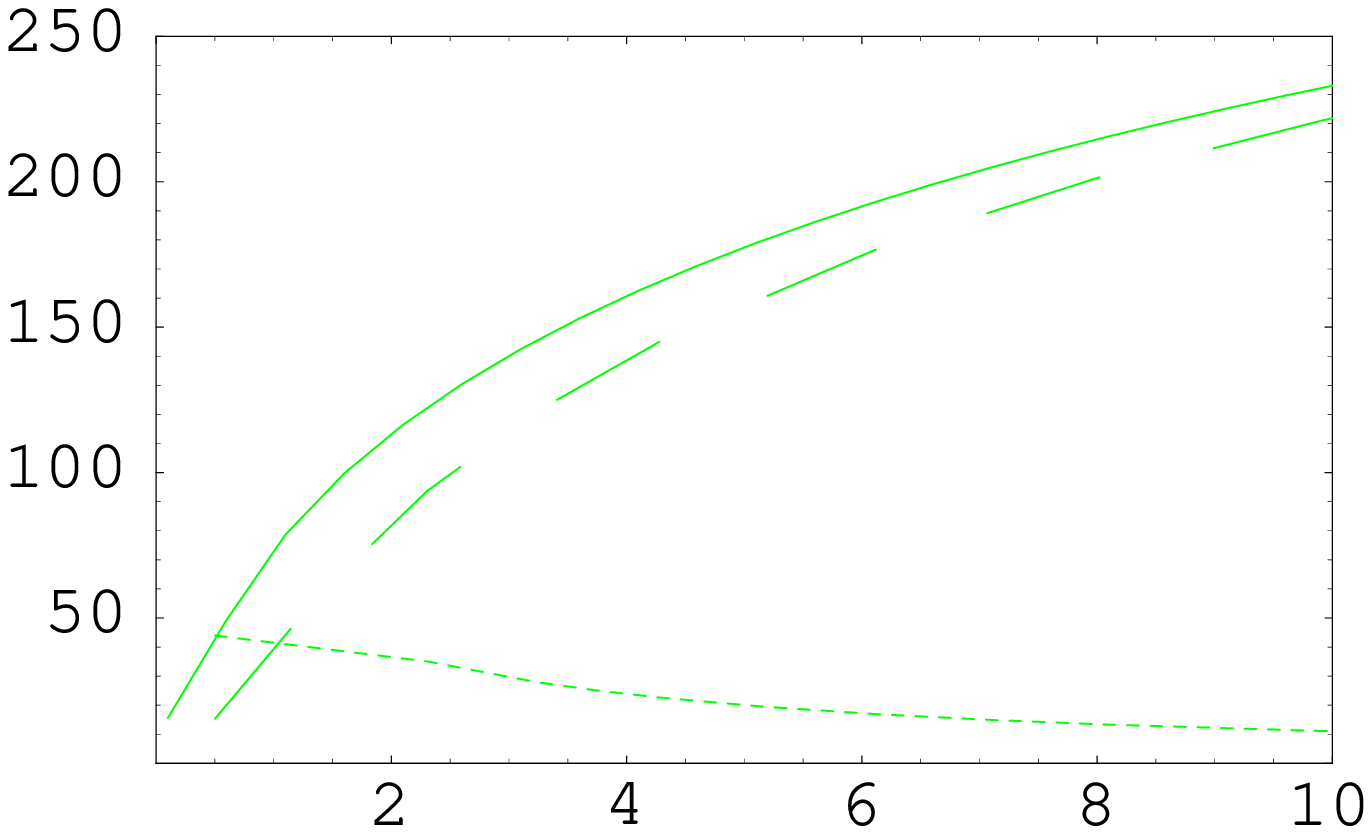,width=50mm,height=50mm}\\

\multicolumn{3}{c}{\rule[-3mm]{0mm}{4mm} $ F_L^D(Q^2)$}\\
$x_B=10^{-2}$&$x_B=5\cdot 10^{-3}$& $x_B=10^{-3}$\\[-10mm]
\epsfig{file=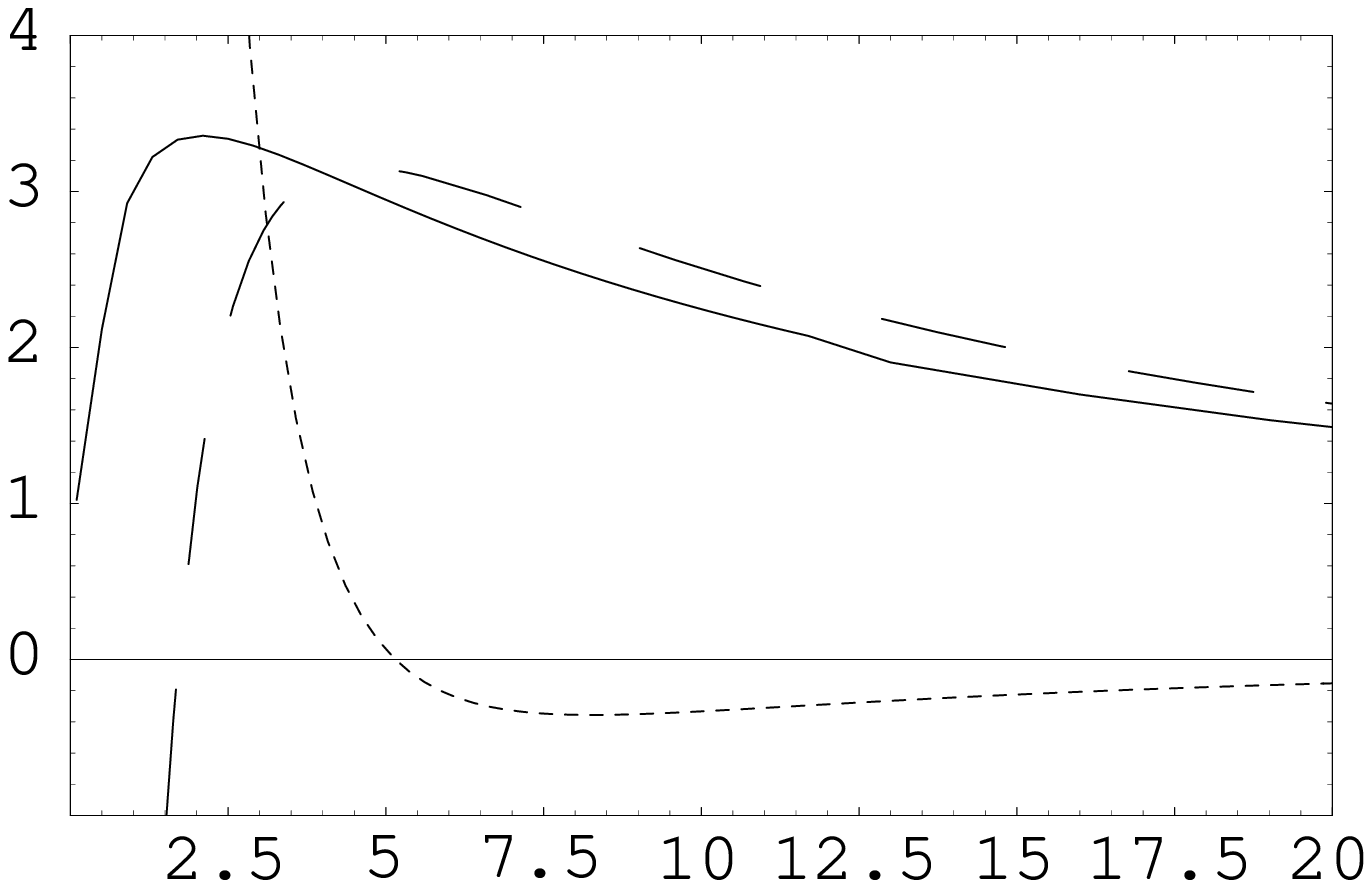,width=50mm,height=50mm} &
\epsfig{file=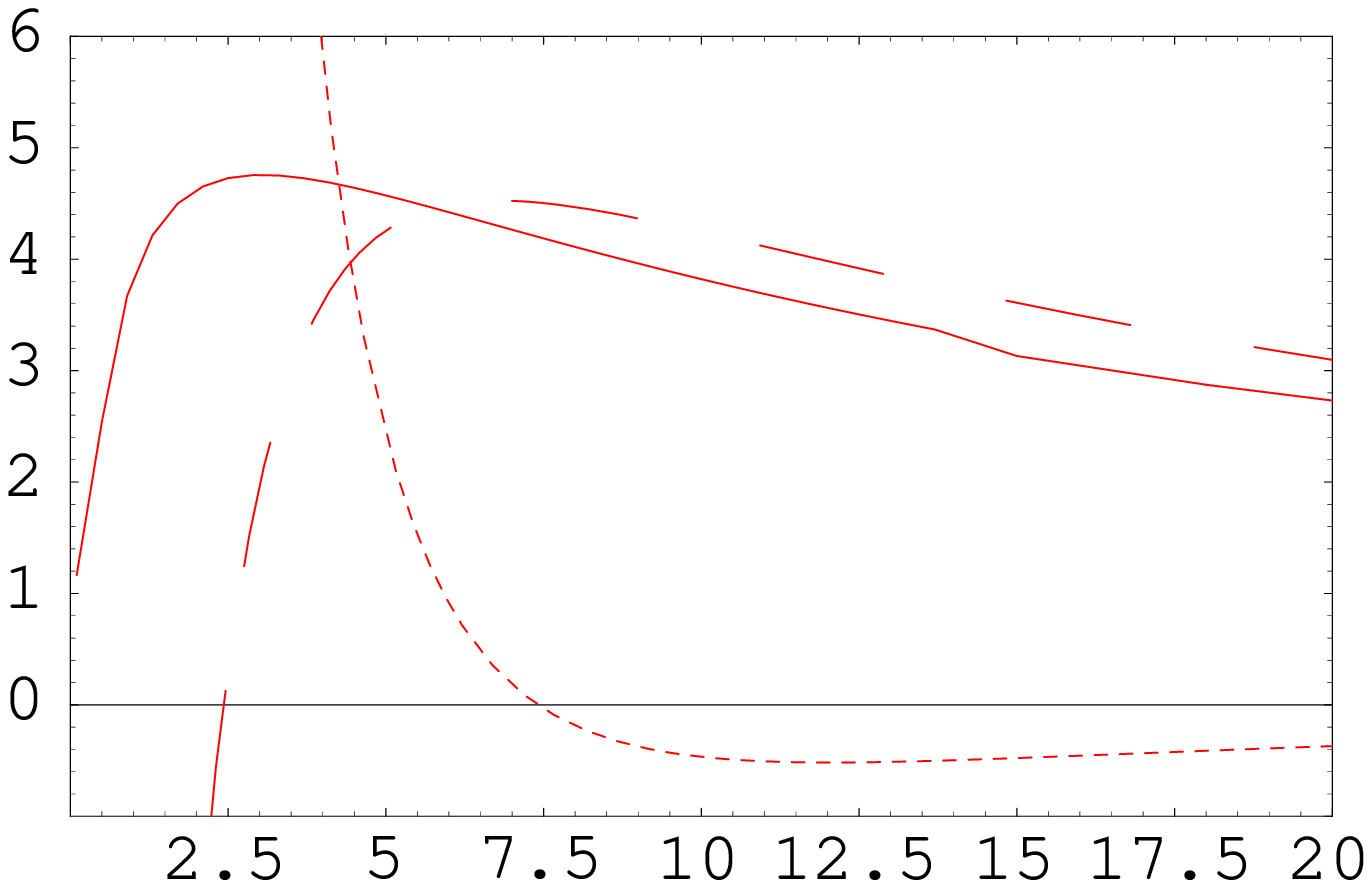,width=50mm,height=50mm} &
\epsfig{file=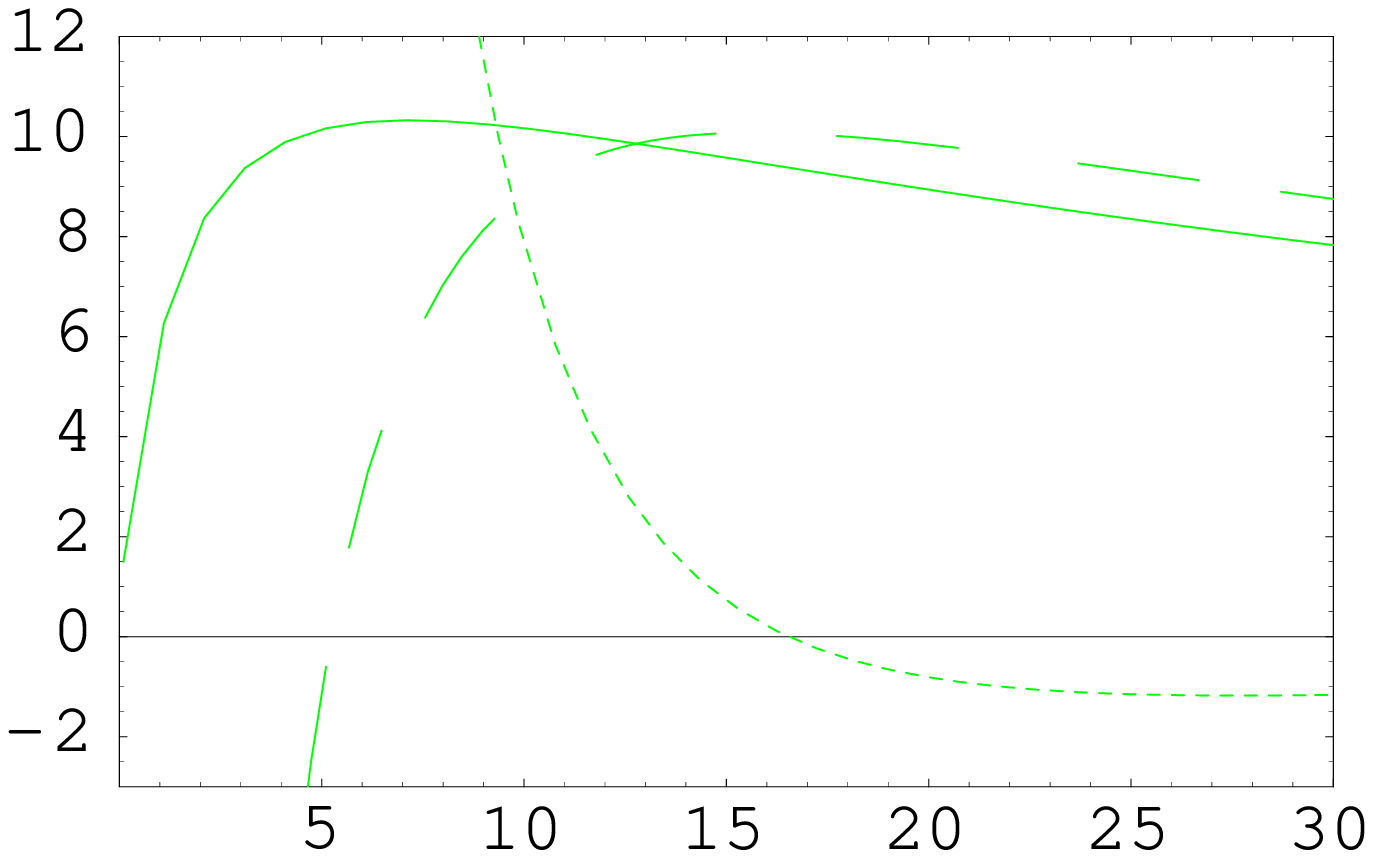,width=50mm,height=50mm}\\

\multicolumn{3}{c}{\rule[-3mm]{0mm}{4mm} $ F_T^D(Q^2)$}\\
$x_B=10^{-2}$&$x_B=5\cdot 10^{-3}$& $x_B=10^{-3}$\\[-10mm]
\epsfig{file=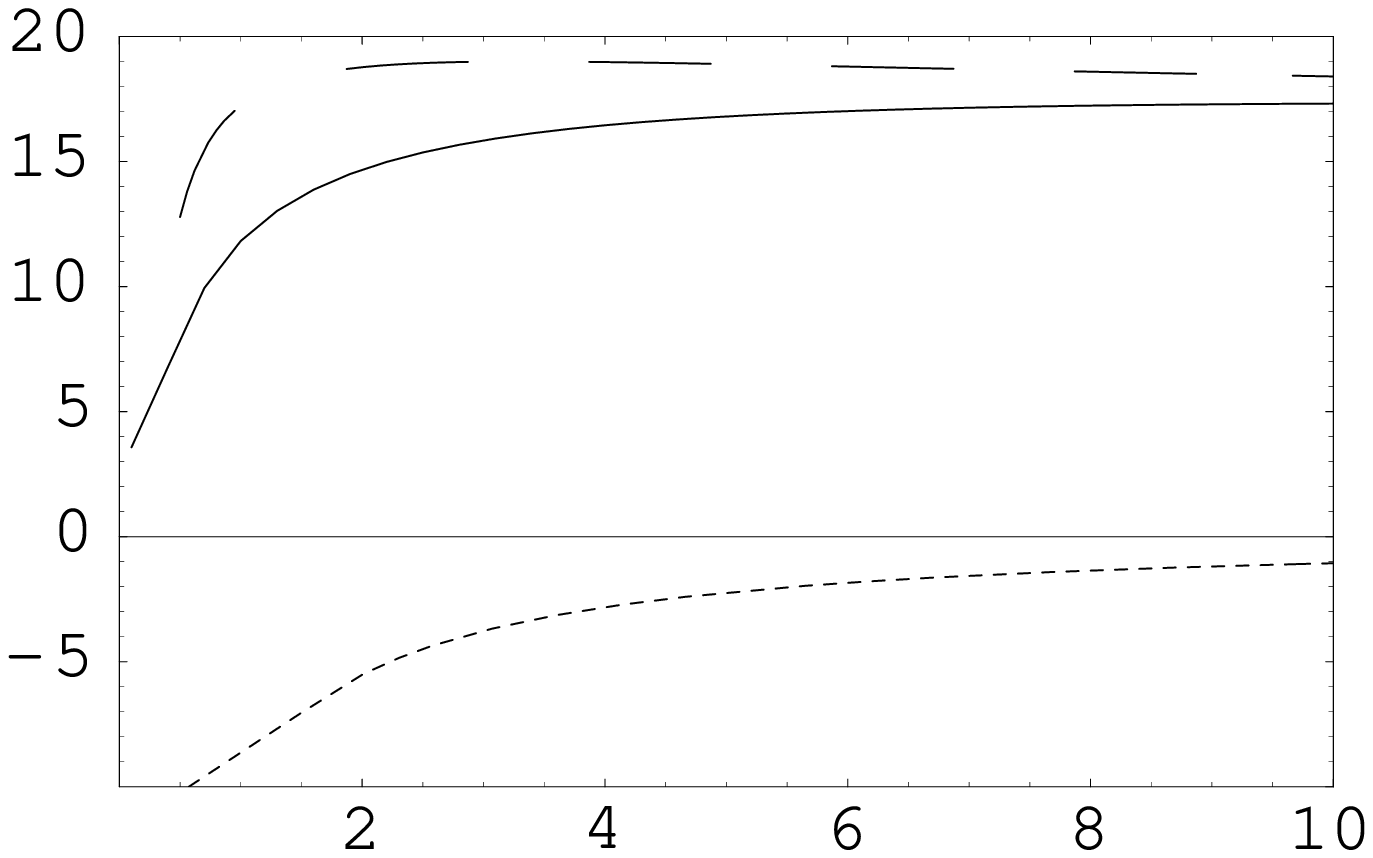,width=50mm,height=50mm} &
\epsfig{file=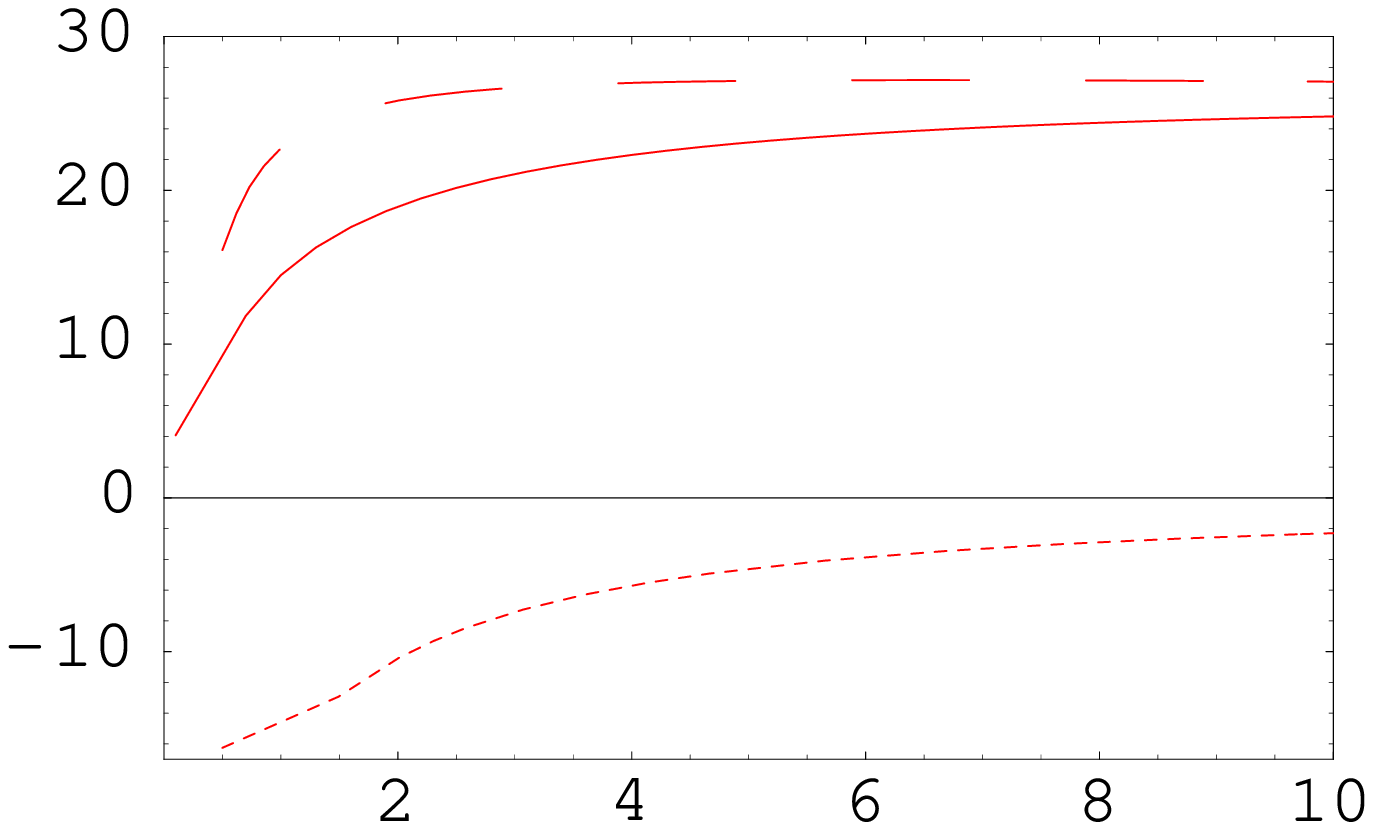,width=50mm,height=50mm} &
\epsfig{file=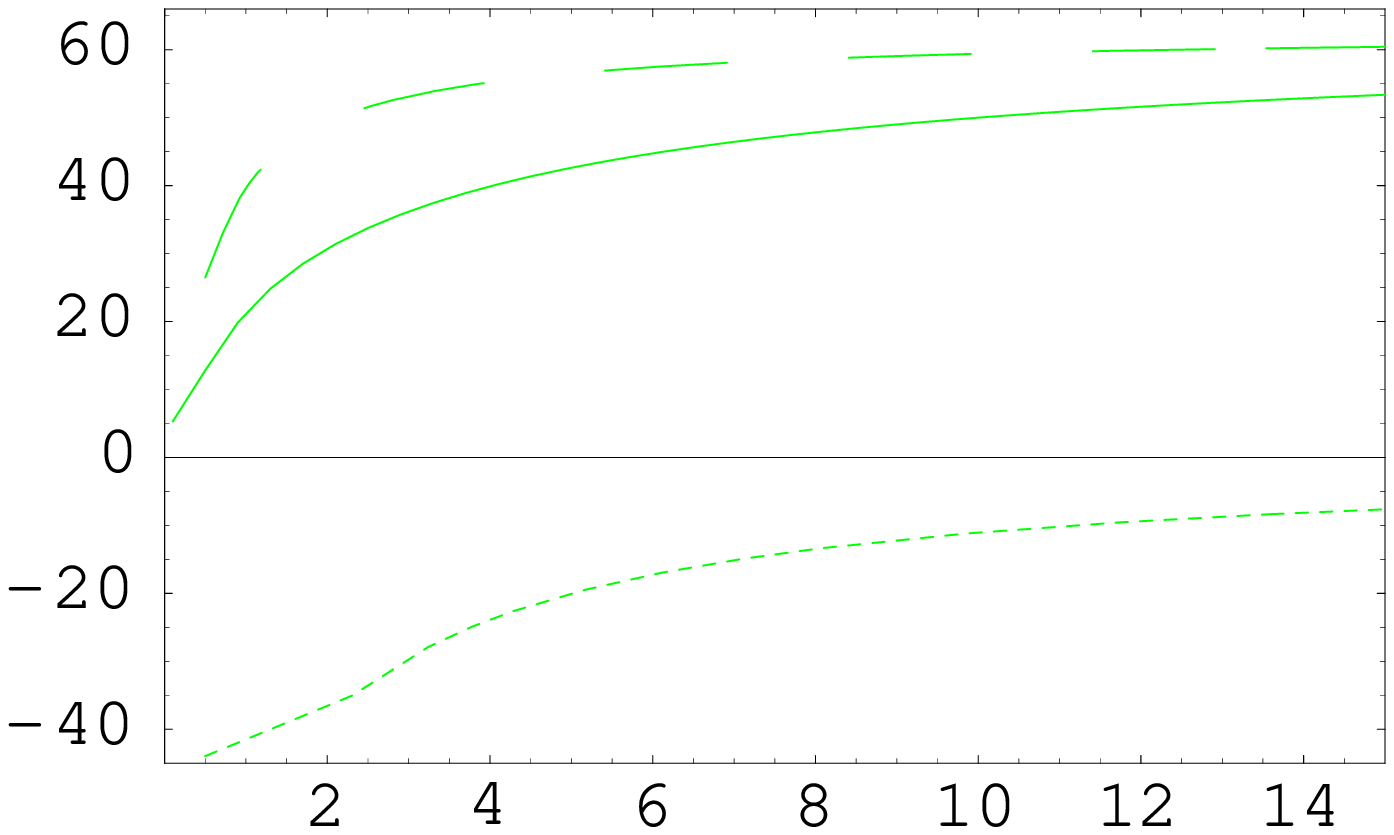,width=50mm,height=50mm}
\end{tabular}
\end{flushleft}
\vspace{-1cm} \caption{\footnotesize {\it Different twists
contributions to the various structure functions for DIS on
nucleus $A$=238 ($U$): leading twist (at high $Q^2$) -- dashed
line, next-to-leading -- dotted one, exact structure function --
solid curve.}}

\label{fig2.238}
\end{figure}

To estimate the contributions of the different twists to various structure
functions 
we perform calculations using \eq{LONG},\eq{TRANS} and \eq{DIFFRACT}.
Note, that \eq{LONG} and \eq{TRANS} are valid only in the  massless
limit $m_f^2=0$. As was explained above, this is a good assumption as long
as we are not concerned with small $Q^2$ behaviour.
All our calculations are made for the eRHIC kinematical region.
The results of the calculations for \emph{Zn} ($A$=30), 
\emph{Sn} ($A$=119) and \emph{U} ($A$=238) are shown in
Figs.~\ref{fig2.30},\ref{fig2.119} and \ref{fig2.238}. 
The following important effects are clearly seen when we compare the
contributions of first two non-vanishing twists:
\begin{enumerate}
\item\quad In all structure functions except $F_L$ there  is a scale $\hat
Q^2$ specific for the particular structure function where these two twists
are
equal. $\hat Q^2$ is largest for $F_L^D$;

\item\quad $\hat Q^2$ grows as $x_B$ decreases and $A$ increases;

\item\quad twist-4 in $F_L$ cancels numerically
with twist-4 in\footnote{This effect was also noted in
Ref.~\cite{BARTELS} where another, though similar, model\cite{WUST}
 was used for SC.} $F_2$  in the region  of $Q^2$ where twist-4 is 
significant (low $Q^2$). 
This is a reason for measuring the polarized structure functions
instead of the total one.

\end{enumerate} 

\begin{figure}
\begin{flushleft}
\begin{tabular}{cc}
$ $& $ $\\[-5mm]
\epsfig{file=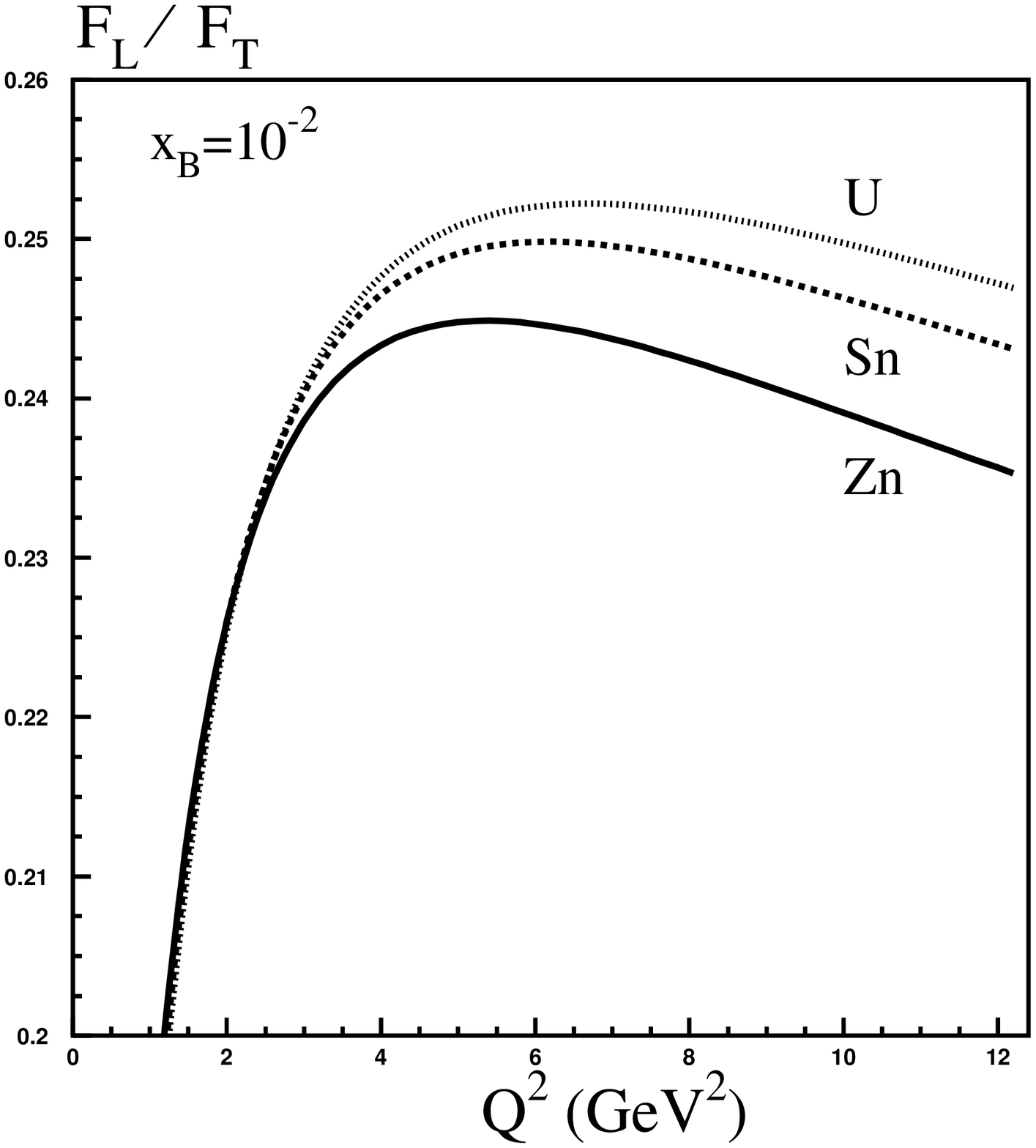,width=80mm,height=80mm}
&
\epsfig{file=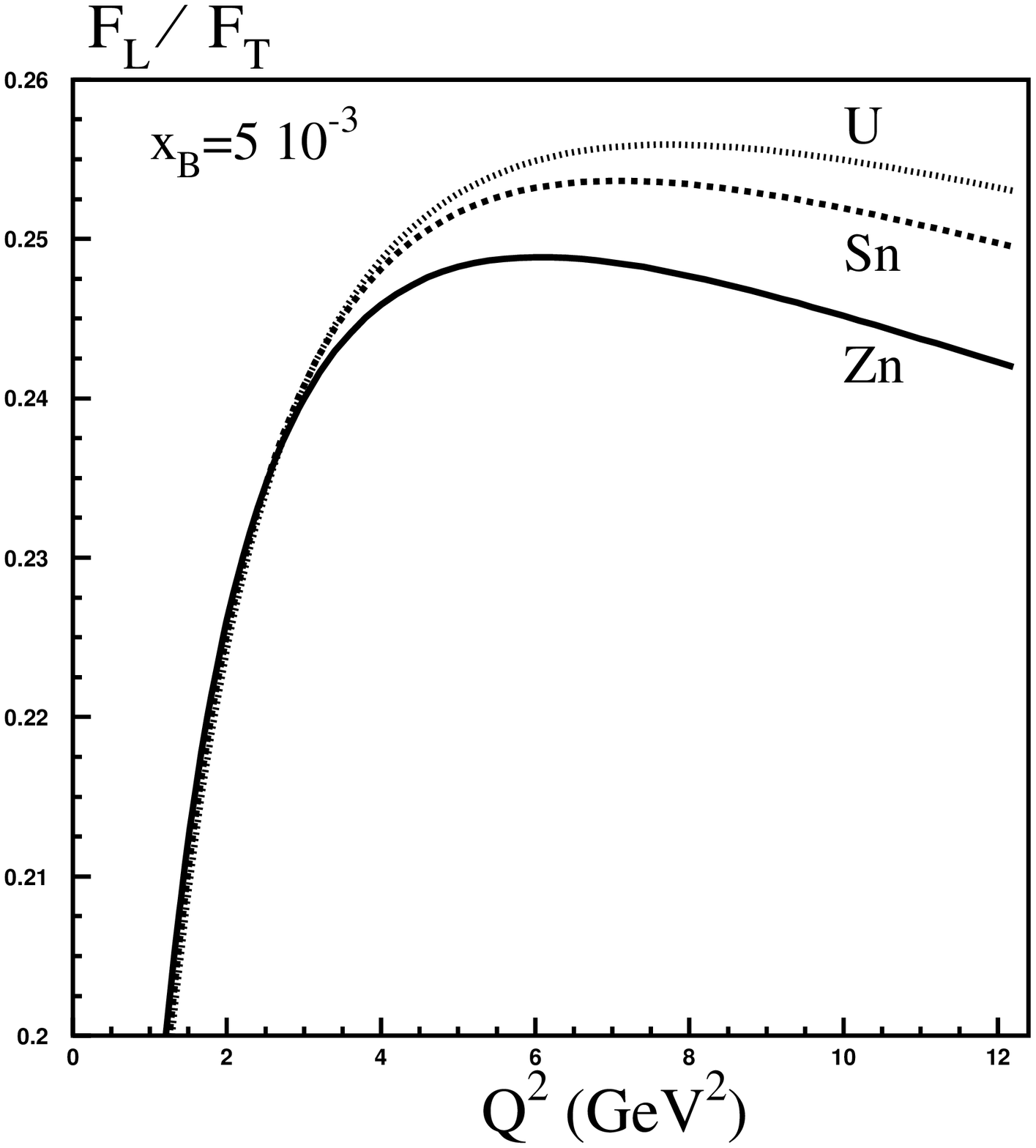,width=80mm,height=80mm}\\
$ $ & $ $\\[-5mm]
\epsfig{file=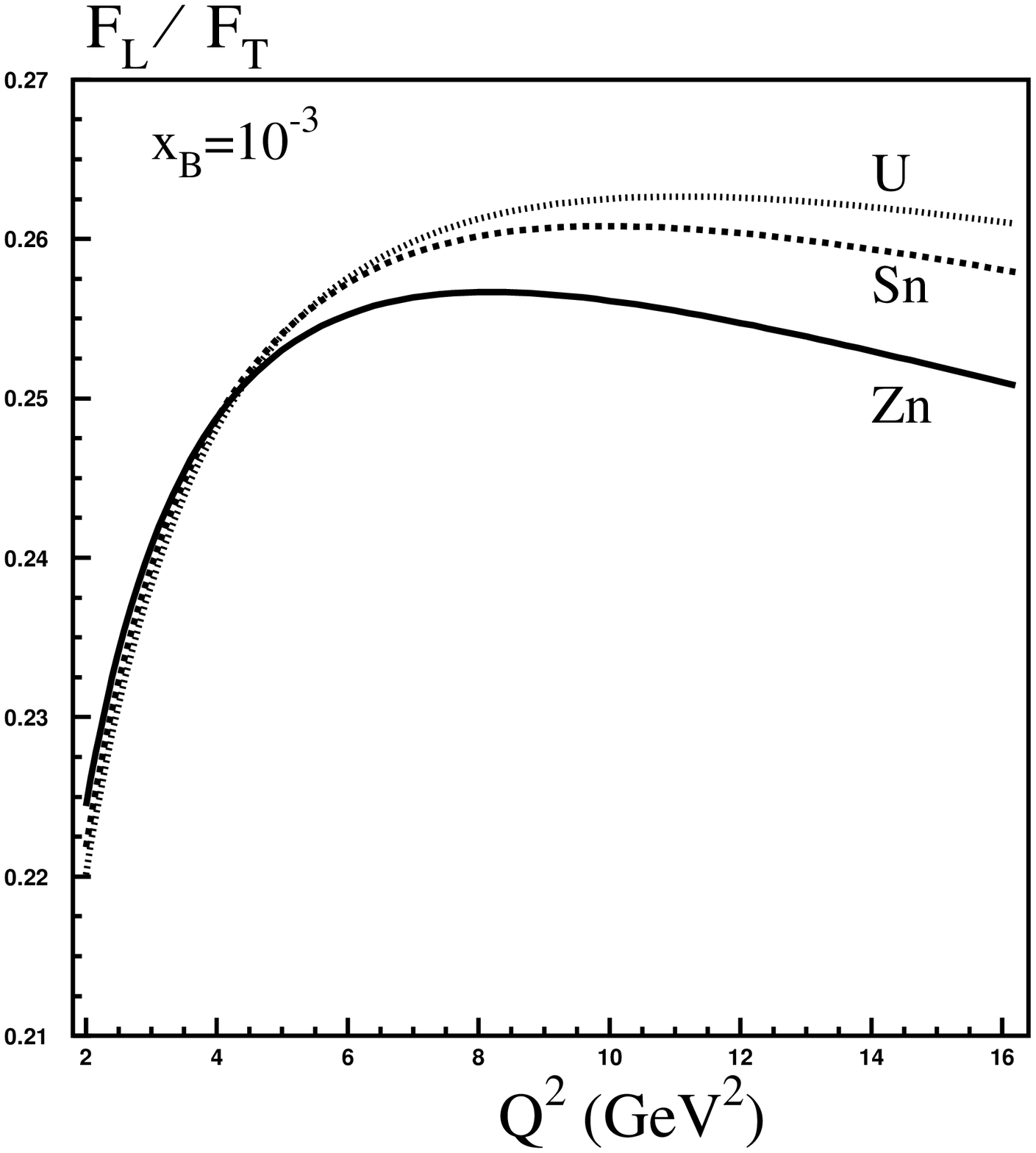,width=80mm,height=80mm}
&
\end{tabular}
\end{flushleft}
\vspace{0.17cm}
\caption{\footnotesize {\it Ratio $F_L/F_T$ versus $Q^2$ for different
$x_B$ and $A$. }}
\label{fig3}
\end{figure}

\begin{figure}
\begin{flushleft}
\begin{tabular}{cc}
$ $& $ $\\[-5mm]
\epsfig{file=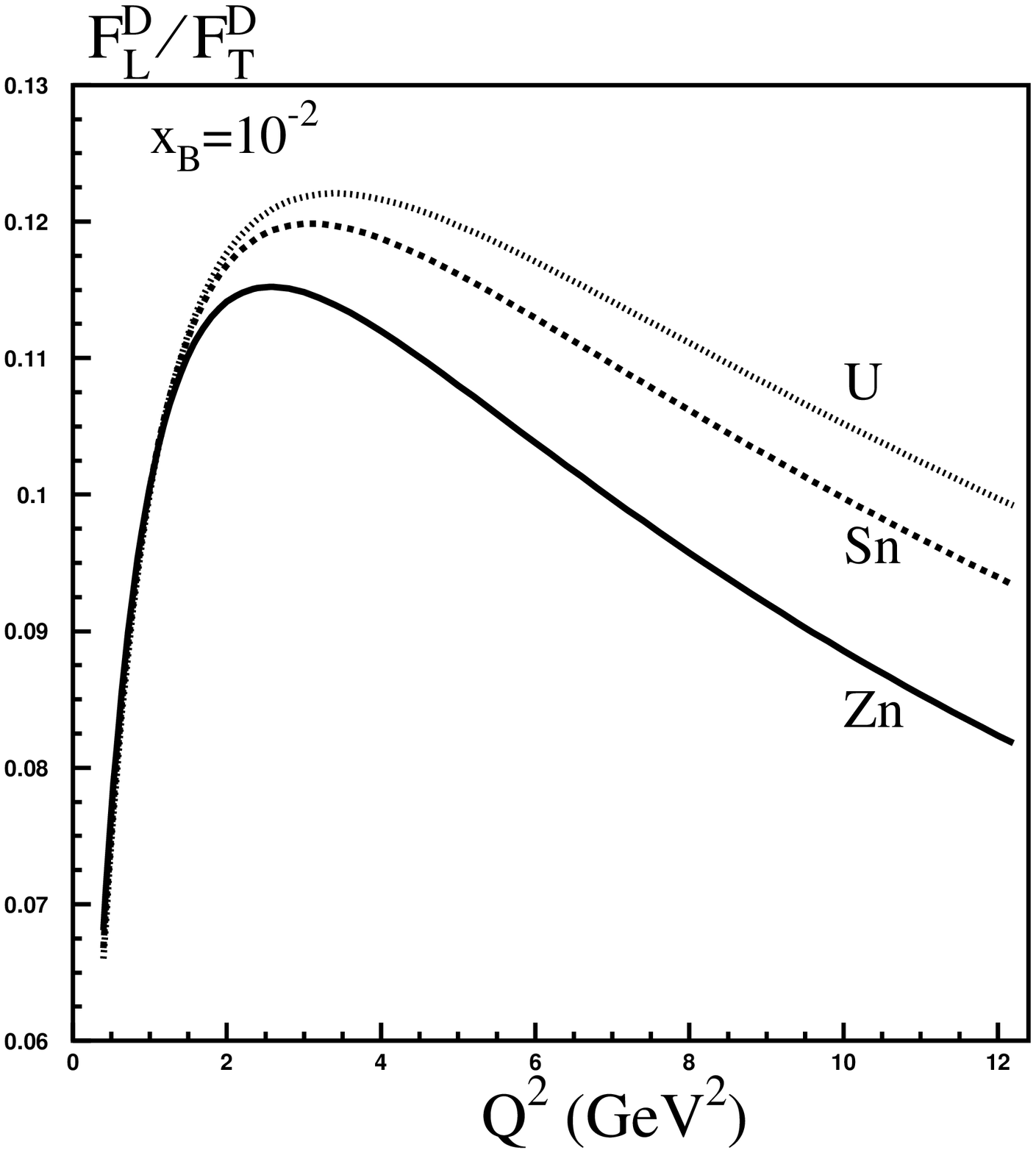,width=80mm,height=80mm}
&
\epsfig{file=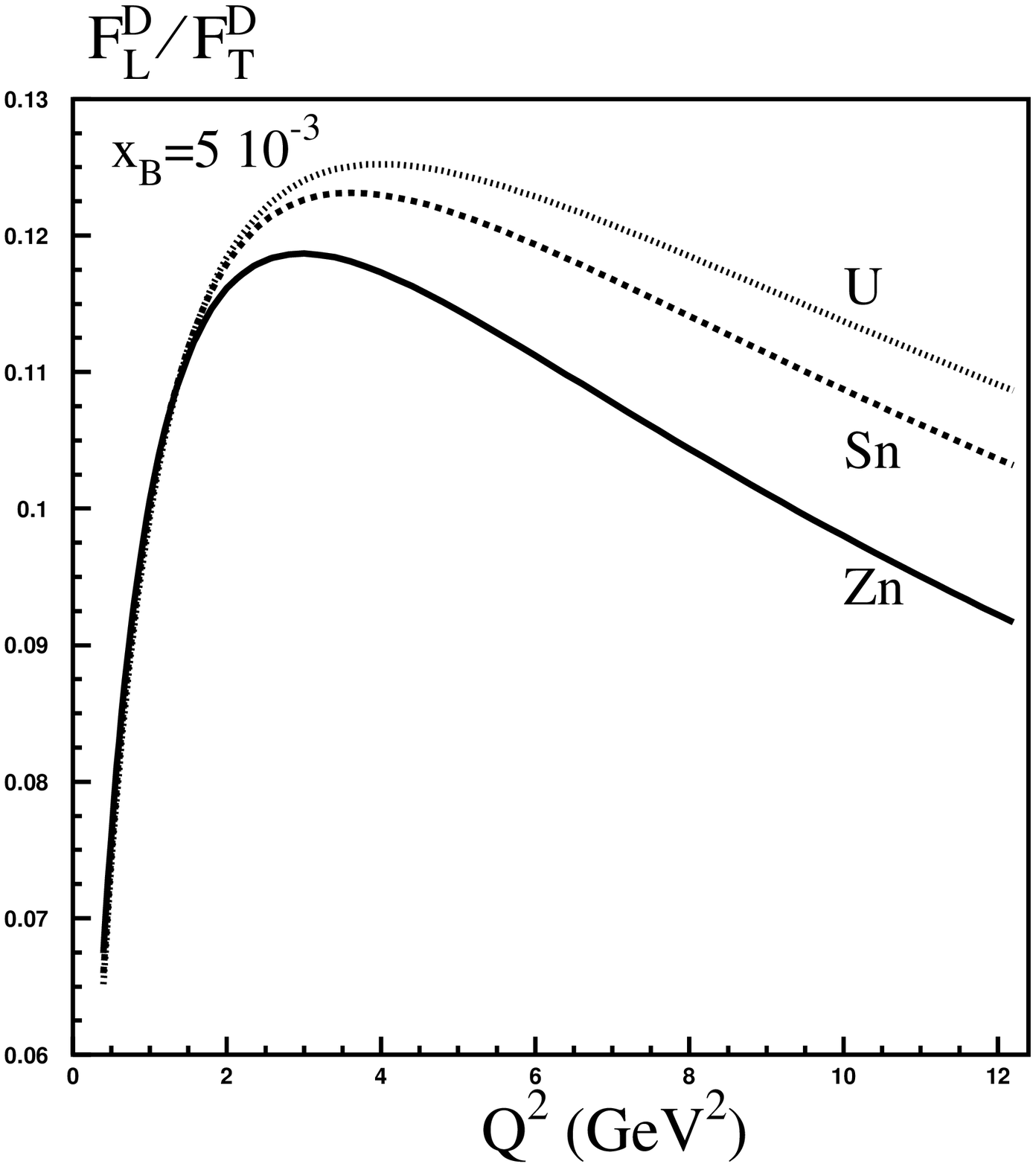,width=80mm,height=80mm}\\
$ $ & $ $\\[-5mm]
\epsfig{file=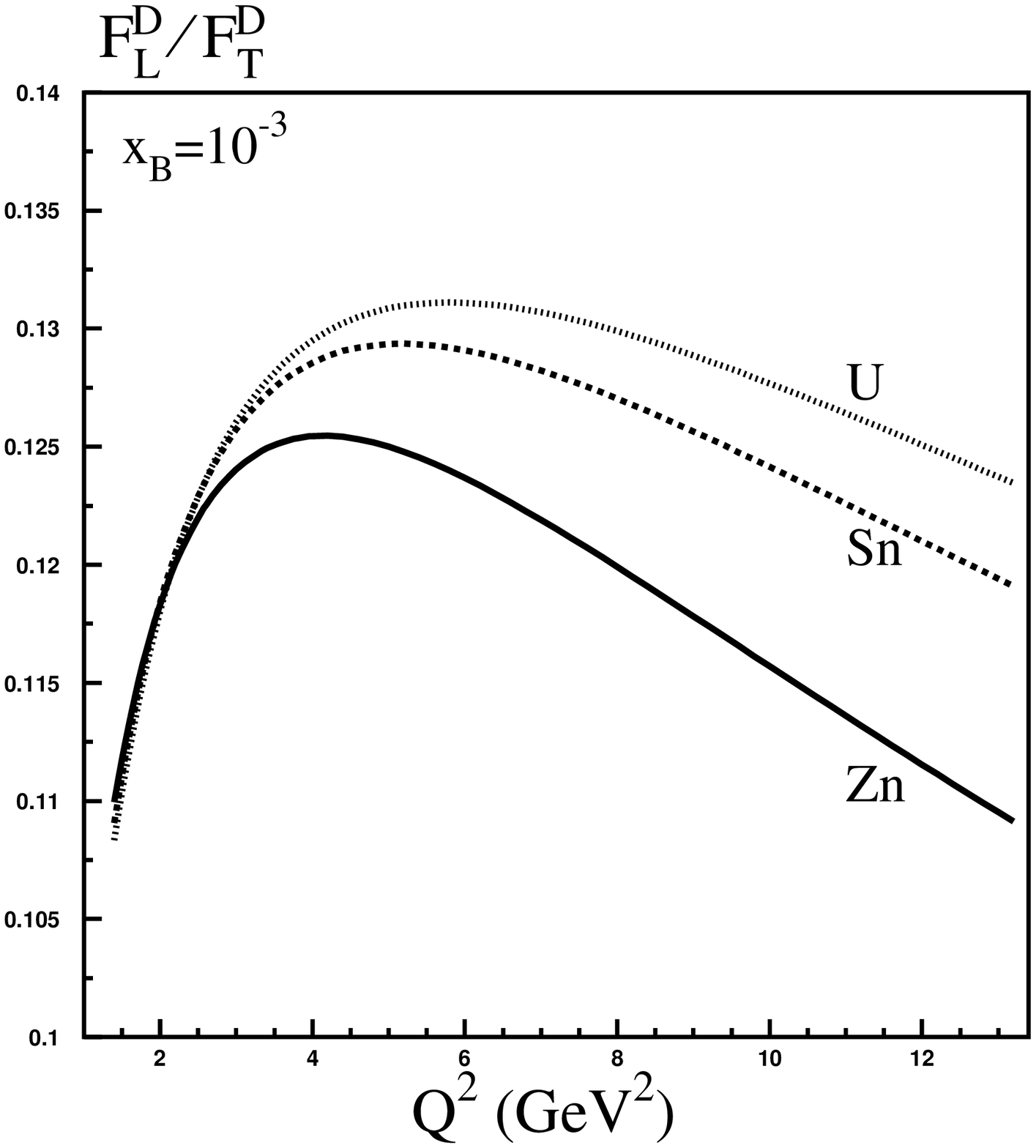,width=80mm,height=80mm}
&
\end{tabular}
\end{flushleft}
\vspace{0.17cm}
\caption{\footnotesize {\it Ratio $F_L^D/F_T^D$ versus $Q^2$ for different    
$x_B$ and $A$. 
}}
\label{fig4}
\end{figure}

In Figs.~\ref{fig3} and \ref{fig4} we plotted the ratios 
$F_L/F_T\,(x_B,Q^2,A)$ and $F_L^D/F_T^D\,(x_B,Q^2,A)$
correspondingly versus $Q^2$. The calculations were performed with 
formulae \eq{TOT} and \eq{DIF} \emph{not} neglecting the mass of the quark
$m_f$.
As is expected on theoretical grounds\cite{LT} the ratio has a
maximum which increases as $x_B$ decreases and $A$ increases.
It occurs at $Q^2_{max}\gg m_f^2$ which implies that we are sufficiently
distant
from the confinement region. However, it differs
 numerically  for inclusive and diffractive ratios. This in
turn implies that the SC will be  still small at eRHIC. 

\begin{figure}
\begin{flushleft}
\begin{tabular}{cc}
$ $& $ $\\[-5mm]
\epsfig{file=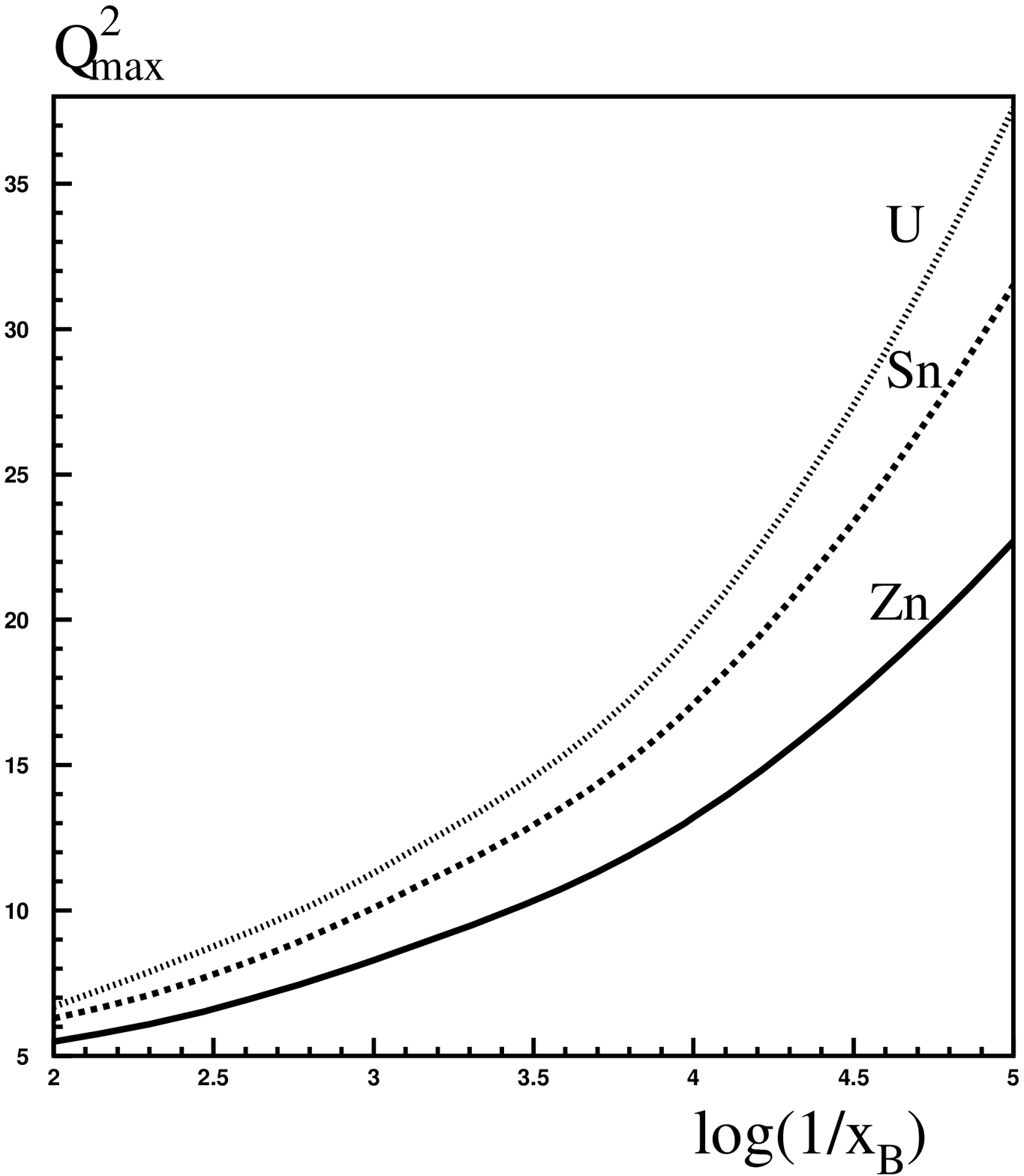,width=80mm,height=80mm}
&
\epsfig{file=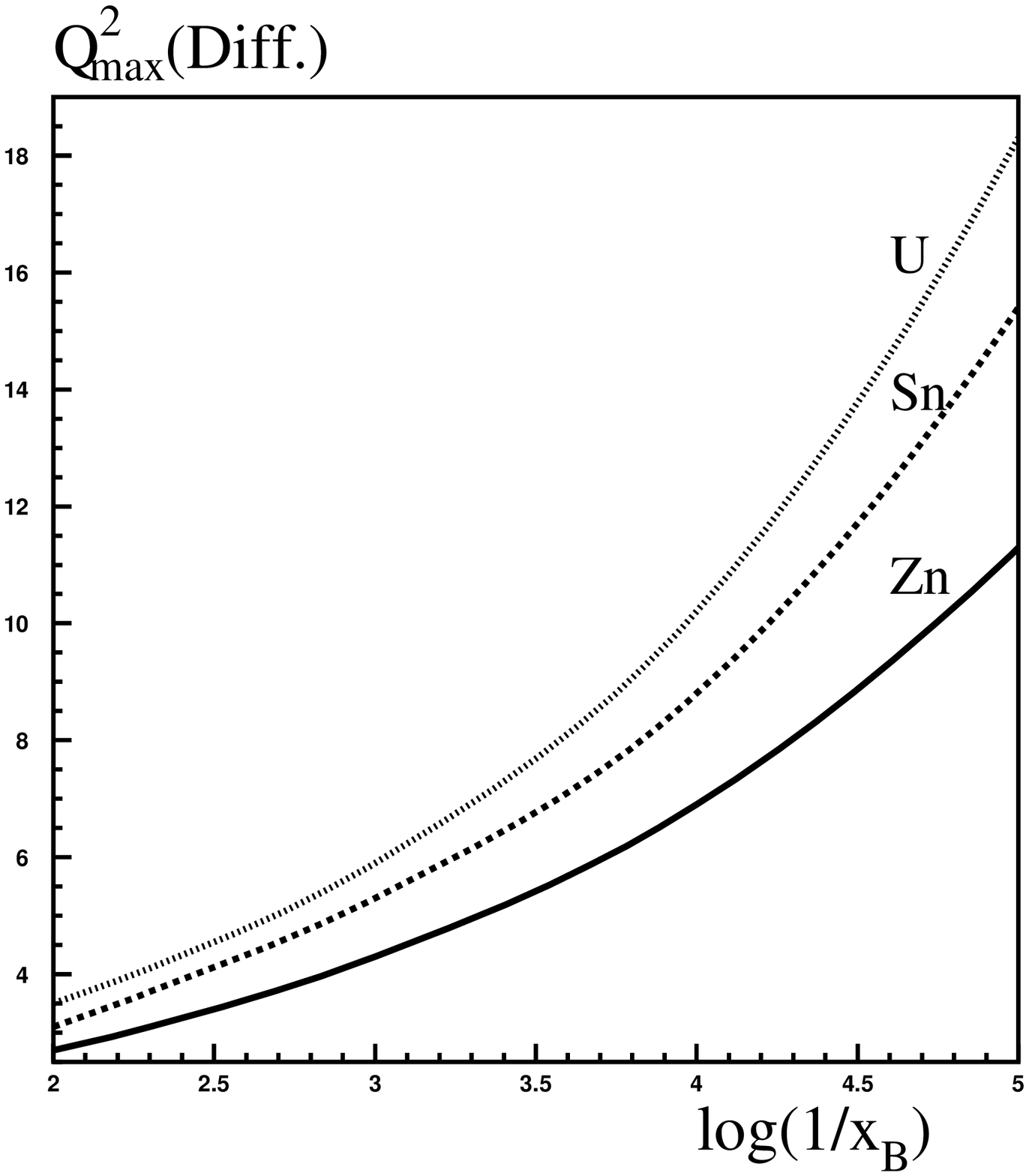,width=80mm,height=80mm}
\end{tabular}
\end{flushleft}
\vspace{0.17cm}
\caption{\footnotesize {\it Scaling of $Q_{max}^2$ with $x_B$:
maxima of ratios of $F_L/F_T$
and $F_L^D/F_T^D$ versus  $\log_{10}(1/x_B)$ for different $A$. }}
\label{fig5}
\end{figure}

\begin{figure}
\begin{flushleft}
\begin{tabular}{cc}
$ $& $ $\\[-5mm]
\epsfig{file=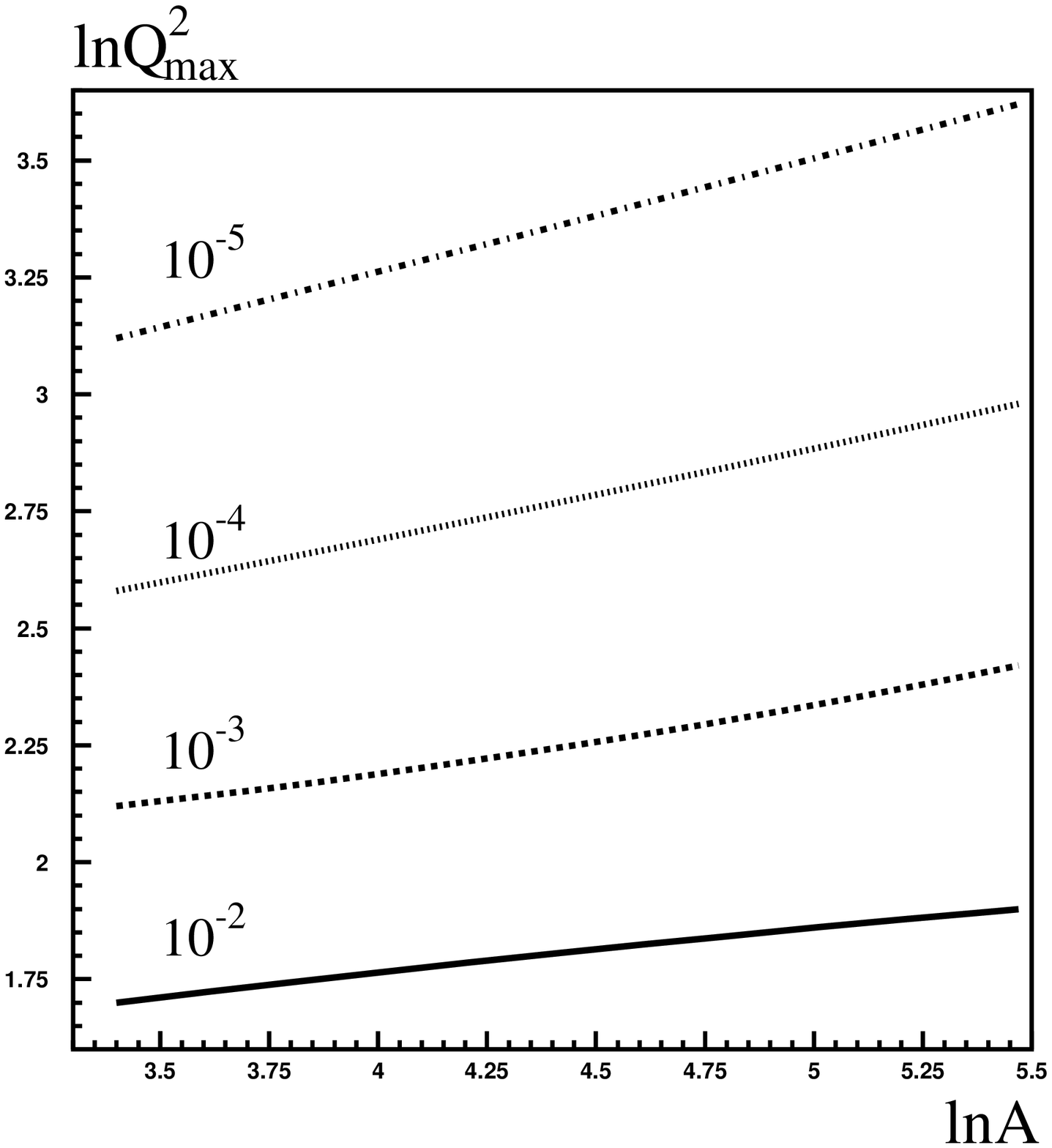,width=80mm,height=80mm}
&
\epsfig{file=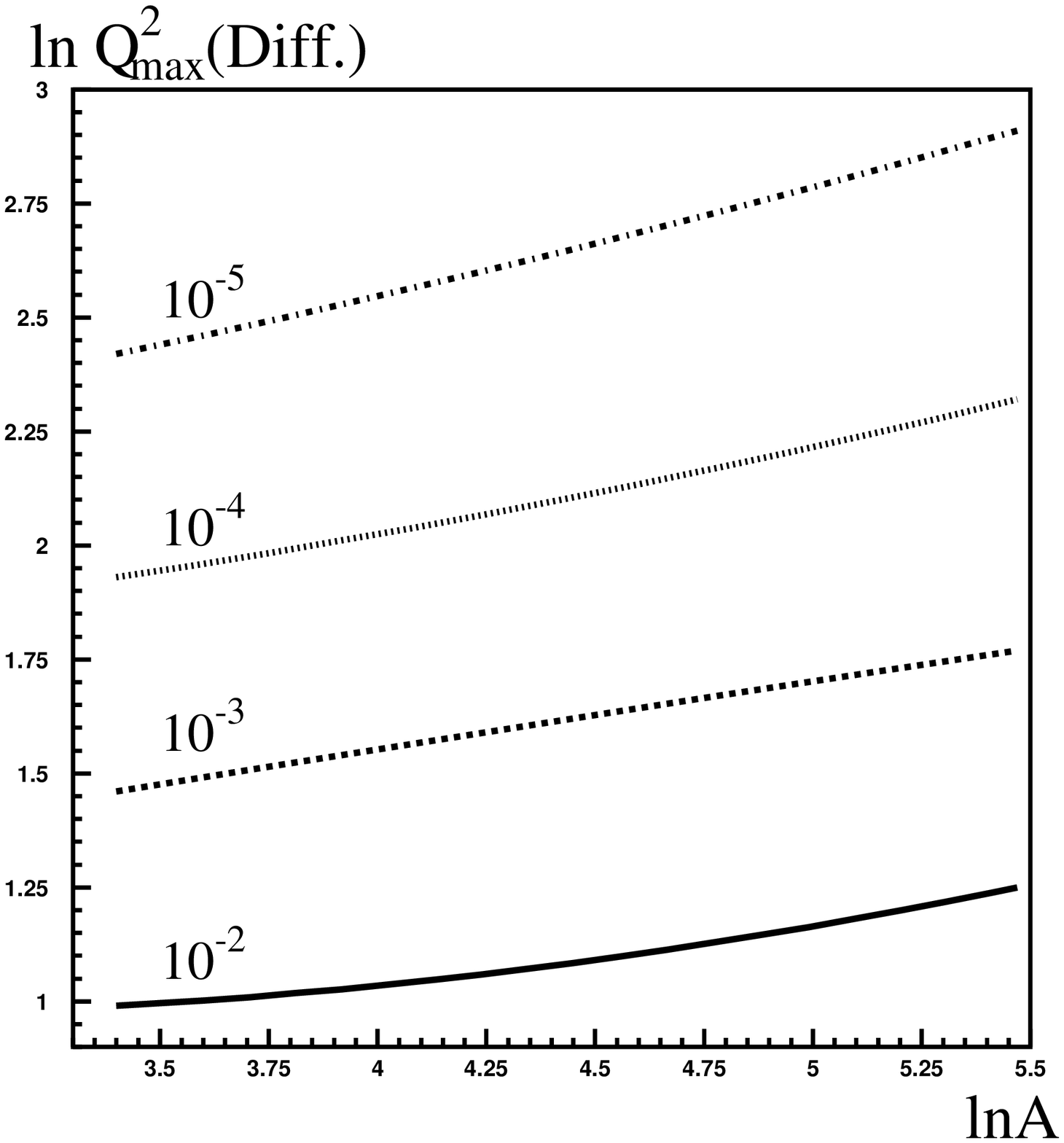,width=80mm,height=80mm}
\end{tabular}
\end{flushleft}
\vspace{0.17cm}
\caption{\footnotesize {\it Scaling of $Q_{max}^2$ with $A$:
$\ln$ of maxima of ratios of $F_L/F_T$ 
and $F_L^D/F_T^D$ versus  $\ln A$ for different $x_B$. }}
\label{fig6}
\end{figure}

In Figs.~\ref{fig5} and \ref{fig6} we show the behavior of
$Q^2_{max}(x_B,A)$
as a function of $x_B$ and $A$. Note in Fig.~\ref{fig6} 
that $Q^2_{max}\sim A^{0.25}$. If the SC were absent  we would
expect that $Q^2_{max}$  not to depend on $A$, while in saturation regime
$Q^2_{max}\sim 
A^{2/3}$. This  confirms once more our conclusion that the kinematical
region of eRHIC is intermediate between the linear and the saturation
regimes.  

\section{Conclusions}
The goal  of our work is to understand to what extent the high parton
density regime of QCD could manifest itself in DIS experiments at
eRHIC where the density of partons could be large enough to make
shadowing corrections  
significant despite the fact that energies are much smaller than that 
achieved at HERA for DIS on nucleons. 
 
The eikonal approach, developed in  Refs. \cite{DIPOLE,BRAZ,DIX}
was used for the first estimates  of the manifestation of 
the saturation scale in deep inelastic scattering with nuclei. It has been
argued that 
it provides a reasonable first iteration to the exact evolution
equation.  We realize that it is not clear whether the 
equations written in the leading $\ln(1/x_B)$ approximation are valid in
the kinematical region of our interest. However, we believe that our
calculations give a reasonable estimate for the magnitude of the effect
and of the kinematic region where this effect could be observed
experimentally.

We observed two manifestations of the saturation scale:
\begin{enumerate}
 \item\,\,\,  The position of
maxima in ratios $F_L/F_T$ and $F_L^D/F_T^D$ ($Q^2_{max}$) is intimely
related to the saturation scale $Q_s^2(x_B)$. $Q^2_{max}$ shows $x_B$ and
$A$
dependence typical for the saturation scale (see Figs. 7 and 8);  
\item\,\,\, 
The results of our calculations show that there exists a scale at which 
the twist expansion breaks down since all twists become of the same order.
We can use the DGLAP evolution equations only for $Q^2$ larger than this
scale. The energy ($x_B$) and $A$ dependence of this scale gives us hope
that the experimental data for deep inelastic scattering with nuclei will
help  to separate leading and higher twists contributions. In the case
of nucleon deep inelastic scattering such a separation appears to be a
rather
difficult task which has not been performed until now. 
\end{enumerate}

With this paper we started a systematic analysis of possible manifestation
of gluon  saturation phenomena in deep inelastic scattering on nuclei.
We hope that our estimates will be useful for planning of future
experiments especially at eRHIC.

\vskip0.3cm
{\large\bf Acknowledgments}
\vskip0.3cm

We thank Jochen Bartels,  Krzysztof Golec-Biernat, Dima Kharzeev and Yura 
Kovchegov for  very fruitful discussions  of saturation phenomena.

This paper in part was supported by  BSF grant  \# 9800276.
The work of L.McL.  was supported by the US Department of Energy
(Contract \# DE-AC02-98CH10886). 
The research of E.L.,  K.T. , E.G. and U.M. was supported in part by the
Israel Science
Foundation, founded by the Israeli Academy of Science and Humanities.


\newpage
%
%

\end{document}